\documentclass[prl,aps,amsmath,amssymb,reprint,superscriptaddress,floatfix,raggedbottom,nofootinbib]{revtex4-2}

\usepackage{graphicx}
\usepackage{dcolumn}
\usepackage{bm}
\usepackage{hyperref}
\usepackage{gensymb}
\usepackage{siunitx}
\usepackage{color}

\newcommand{\ignore}[1]{}

\begin{document}

\title{Self-Kerr effect across the yellow $\mathrm{Cu_2O}$ Rydberg series}

\author{Corentin Morin}
\affiliation{Laboratoire de Physique de l'\'Ecole Normale Sup\'{e}rieure (LPENS), ENS-Universit\'{e} PSL, Sorbonne Universit\'{e}, CNRS, Paris 75005, France}

\author{J\'er\^ome Tignon}
\affiliation{Laboratoire de Physique de l'\'Ecole Normale Sup\'{e}rieure (LPENS), ENS-Universit\'{e} PSL, Sorbonne Universit\'{e}, CNRS, Paris 75005, France}

\author{Juliette Mangeney}
\affiliation{Laboratoire de Physique de l'\'Ecole Normale Sup\'{e}rieure (LPENS), ENS-Universit\'{e} PSL, Sorbonne Universit\'{e}, CNRS, Paris 75005, France}

\author{Sukhdeep Dhillon}
\affiliation{Laboratoire de Physique de l'\'Ecole Normale Sup\'{e}rieure (LPENS), ENS-Universit\'{e} PSL, Sorbonne Universit\'{e}, CNRS, Paris 75005, France}

\author{Gerard Czajkowski}
\affiliation{Institute of Mathematics and Physics, Technical University of Bydgoszcz, Poland}

\author{Karol Karpiński}
\affiliation{Institute of Mathematics and Physics, Technical University of Bydgoszcz, Poland}

\author{Sylwia Zielińska-Raczyńska}
\affiliation{Institute of Mathematics and Physics, Technical University of Bydgoszcz, Poland}

\author{David Ziemkiewicz}
\affiliation{Institute of Mathematics and Physics, Technical University of Bydgoszcz, Poland}

\author{Thomas Boulier}\email{thomas.boulier@phys.ens.fr}
\affiliation{Laboratoire de Physique de l'\'Ecole Normale Sup\'{e}rieure (LPENS), ENS-Universit\'{e} PSL, Sorbonne Universit\'{e}, CNRS, Paris 75005, France}

\date{\today}

\begin{abstract}
We investigate the nonlinear refraction induced by Rydberg excitons in $\mathrm{Cu_2O}$. Using a high-precision interferometry imaging technique that spatially resolves the nonlinear phase shift, we observe significant shifts at extremely low laser intensity near each exciton resonance. From this, we derive the nonlinear index $\mathrm{n_2}$, present the $\mathrm{n_2}$ spectrum for $n\geq 5$ and report large $\mathrm{n_2}$ values of order $\si{10^{-3}\milli\meter^2/\milli\watt}$. Moreover, we observe a rapid saturation of the Kerr nonlinearity and find that the saturation intensity $\mathrm{I_{sat}}$ decreases as $n^{-7}$. We explain this with the Rydberg blockade mechanism, whereby giant Rydberg interactions limit the exciton density, resulting in a maximum phase shift of 0.5 rad in our setup.

\end{abstract}
\maketitle

Nonlinear optics has been of paramount scientific importance, unlocking an impressive number of technologies now used in areas ranging from telecommunication and data storage to quantum research. While early studies focused on inorganic crystals as a nonlinear medium, nowadays record nonlinearities are obtained from the coherent manipulation of atomic resonances~\cite{fleischhauer2005electromagnetically}. In particular, dense ultra-cold atomic gases excited to a high principal quantum number $n$ (a Rydberg state) can induce strong nonlinearities at the level of individual photons~\cite{gorshkov2011photon,peyronel2012quantum}. Following in the footsteps of their atomic cousins, Rydberg excitons are attracting considerable attention as they represent an enticing path towards more scalable solid-state Rydberg systems~\cite{assmann2020semiconductor}, with potential for quantum simulation~\cite{taylor2021simulation} and photon logic~\cite{walther2018giant,walther2021nonclassical}. 

In copper oxide ($\mathrm{Cu_2O}$), the semiconductor where excitons were first discovered, principal quantum numbers up to $n=30$ were observed~\cite{kazimierczuk2014giant,heckotter2020experimental,versteegh2021giant}. These correspond to gigantic electron-hole wavefuntions that can span several microns in diameter and were confirmed to have the same $n$-scaling as atoms~\cite{heckotter2017scaling}. In particular, the  $n^{-3}$ scaling of linewidths rapidly leads to sharp resonances while the $n^{11}$ scaling for van der Waals interactions generates a strongly nonlinear absorption due to the phenomenon of Rydberg blockade~\cite{kazimierczuk2014giant}. Since signatures of coherence were observed~\cite{grunwald2016signatures}, several theoretical studies focused on the optical nonlinearities of Rydberg excitons~\cite{zielinska2019nonlinear,walther2020controlling,walther2020plasma} in view of their potential for nonlinear quantum optics~\cite{khazali2017single,walther2018giant,ziemkiewicz2020electromagnetically,walther2021nonclassical}. Experimental studies on Rydberg exciton nonlinearities are dominated by second harmonic generation (SHG) as a potent spectroscopic tool~\cite{farenbruch2020rydberg,rogers2021high,mund2018high,farenbruch2021second} and by investigating the blockade-induced nonlinear absorption~\cite{kazimierczuk2014giant,heckotter2021asymmetric, gallagher2022microwave}. However, so far, no experimental study has looked at the giant Kerr-type optical nonlinearities expected from the sharp Rydberg resonances~\cite{zielinska2019nonlinear}. This is in spite of their important role for nonlinear quantum optics, as Kerr nonlinearities are equivalent to photon-photon interactions and have been instrumental both for applications (e.g. Kerr mode-locking~\cite{brabec1992kerr}) and fundamental investigations (e.g. superfluids of light~\cite{carusotto2013quantum,fontaine2018observation,boulier2020microcavity}, nonlinear photonics~\cite{peyronel2012quantum,chang2014quantum}). Indeed, a Kerr nonlinearity operating at the scale of a few photons is key to all-optical quantum information processing, a paramount goal in the current context where photons are major quantum information conveyors. Thus, a condensed matter medium supporting giant nonlinear indices is of strong interest, and in this context Rydberg excitons were proposed as a more scalable alternative to ultra-cold atomic gases~\cite{walther2018giant,ziemkiewicz2020electromagnetically}. Therefore, to map the potential of Rydberg excitons for nonlinear optics, their ability to generate a large Kerr coefficient must be explored.

To this end, we reveal the giant nonlinear optical index caused by the sharp Rydberg resonances in a resonant one-photon experiment and observe a Kerr coefficient up to $10^{14}$ larger than in typical nonlinear crystals. Moreover, we also observe a rapid saturation of the optical nonlinearity at low power that originates from Rydberg interactions. Including the Rydberg blockade in our model yields an excellent agreement with the experiment. The experimental method is an improved, high-precision variation of interferometric phase front imaging~\cite{olbright1986interferometric} able to accurately map small Kerr phase shifts. The present investigation of the nonlinear refraction in $\mathrm{Cu_2O}$ therefore provides a complimentary insight to the previous nonlinear absorption studies. 

\begin{figure*}[t]
\begin{center}
  \includegraphics[width=1.0\linewidth]{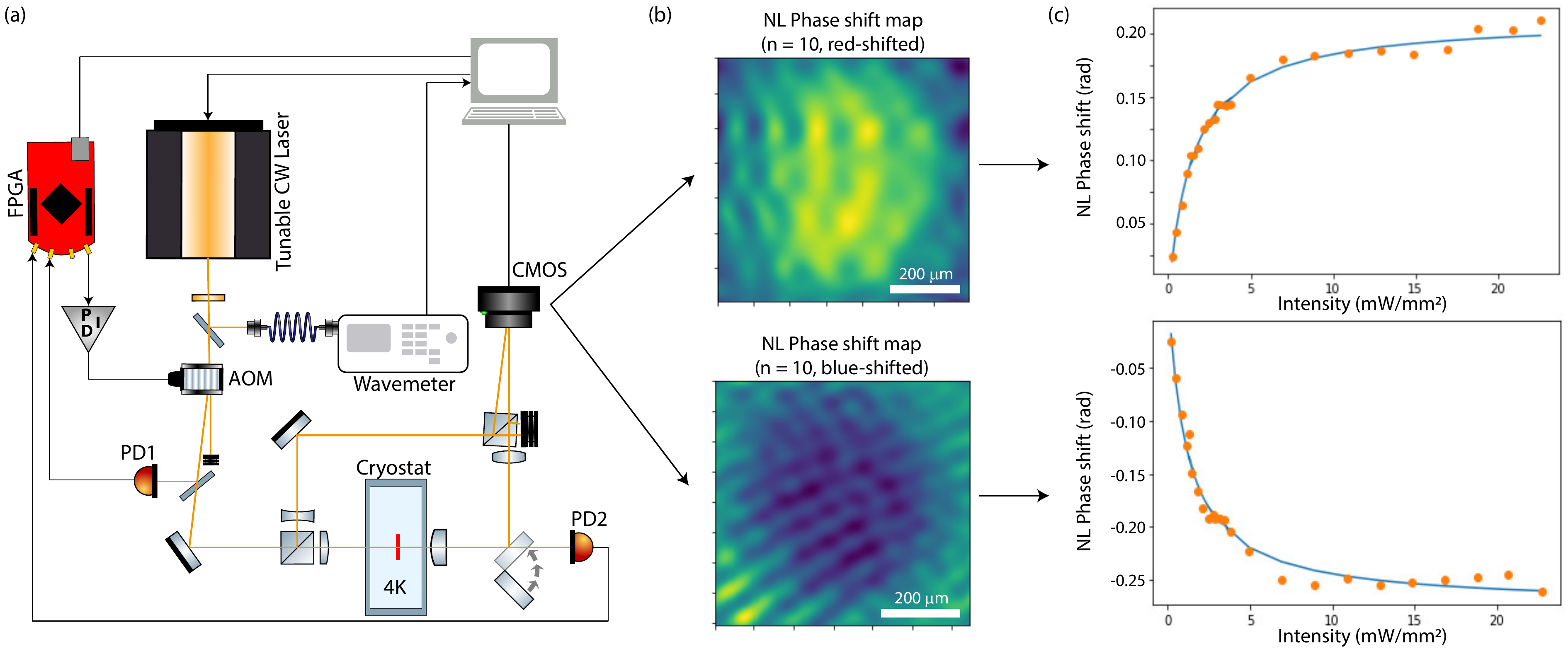}
  \caption{(a) Simplified view of the experimental apparatus. The main computer controls all elements so that data-taking is fully automated. A motorized flipper mirror switches between the transmission measurement and the imaging system. (b) Examples of phase profiles reconstructed from interferograms: red-detuned (up) and blue-detuned (down) from the n = 10 resonance. The phase shift pattern matches the input intensity profile (not shown here). (c) Intensity dependence of the phase shift, extracted from the intensity and phase pictures for each laser energy. We fit the data with a saturable function to extract the initial slope and the saturation intensity.}
  \label{fig1}
\end{center}
\vspace{-5mm}
\end{figure*}

Figure~\ref{fig1} (a) is a simplified representation of the experiment. The sample is a $50~\si{\micro\meter}$ thick natural $\mathrm{Cu_2O}$ crystal, highly polished on both sides and oriented so that light propagates along the [001] axis. It is held strain-free in a $4~\si{\kelvin}$ cryostat in transmission mode. A CW yellow laser (tunable doubled OPO system) provides a spectrally narrow beam, focused on the sample with a waist diameter of about $400~\si{\micro\meter}$. Both the laser frequency and intensity are locked, allowing $\pm 30~\si{\mega\hertz}$ frequency accuracy and reducing the intensity fluctuations to $\lesssim 1\%$ of the mean value. The latter is important for our highly nonlinear system and allows to modulate the laser beam to avoid heating. The transmission is measured during a single $10~\si{\milli\second}$ pulse by a pair of amplified photodiodes. A motorized flip mirror switches between the photodiode transmission measurement and the imaging system used to spatially resolve the transmitted intensity and phase. Thus, we independently measure the optical absorption and refraction in the crystal. The main computer orchestrates a fully-automated measurement protocol, including quality assessment routines.

\ignore{\textit{Phase front imaging} - }We image the phase front of the beam transmitted through the sample using a modified Mach–Zehnder interferometer. One arm is focused on the sample plane and imaged on a camera, while the other is expanded with a divergent lens and used as a quasi-flat reference. The sample image and the reference arrive on the CMOS chip at an angle so that high-contrast fringes containing the phase profile information are present in the recorded picture. The actual phase picture is then retrieved numerically with a Fourier transform algorithm (described in the Supplemental). Shutters allow to image both the interference pattern and the intensity profile. Moreover, for each laser energy we typically take two sets of pictures: one at the desired laser power $P$ and one at vanishing laser power (at least $50\times$ lower than $P$). The low power data contains a negligible amount of nonlinear phase shift compared with the high power data. We therefore subtract the low power phase profile from the high power phase profile to ensure that only intensity-dependant effects remain. The resulting nonlinear phase shift mapping is exemplified in figure~\ref{fig1} (b). Such subtraction of the linear phase profile has the advantage of neatly removing systematic imperfections (e.g. the residual parabolic phase front of the reference beam and small optical aberrations) and was found crucial to reach a sufficient resolution for our $\mathrm{Cu_2O}$ system. Another important noise source is the air fluctuations, which we reduced by carefully shielding our small interferometer and by integrating each image for longer than the dominant fluctuation timescale, found to be $\lesssim 100~\si{\milli\second}$. The typical phase resolution of our setup is of order $\pm 0.01$ rad.

\ignore{\textit{Nonlinear optical response} - }The Gaussian profile of the input beam contains all intensities between zero and its maximum $I_{max} = P/(2\pi\sigma^2)$, where $\sigma=200~\si{\micro\meter}$ is the Gaussian radius. Therefore, the self-Kerr effect induces a non-uniform nonlinear phase shift pattern roughly resembling the input intensity profile (see figure~\ref{fig1} (b)). We extract the phase shift dependence on the input intensity $\Delta\phi(I)$ from the intensity and phase images by calculating the average phase shift for sets of pixels containing the same intensity. Figure~\ref{fig1} (c) presents examples of $\Delta\phi(I)$ for two different energies, red- and blue-detuned from the $n = 10$ exciton resonance. The expected change of sign of the Kerr coefficient is well visible. We report here the maximum observed phase shift, the nonlinear optical index $\mathrm{n_2}$ deduced from the initial slope $\frac{\partial\Delta\phi}{\partial I}|_{I \to 0}$ and the saturation intensity $I_{sat}$ typically observed near exciton resonances.

This approach has significant advantages over the traditional z-scan technique~\cite{sheik1989high}. It does not need a perfect Gaussian beam because we spatially resolve both the intensity and phase signals. It is therefore robust to optical aberrations. Additionally, our approach does not require moving parts: it is faster, more stable and, importantly, compatible with complex setups that cannot be moved, such as cryostats (as is the case here). It is typically more precise and it is cost-effective: the most expansive part is the camera, which does not need to be especially fast or sensitive. Moreover, we directly access the phase shift $\Delta\phi(I)$: the measured quantity is physically relevant and the fitting is simple, revealing both $\mathrm{n_2}$ and $I_{sat}$. Finally, unlike with the z-scan technique, focusing the laser is not necessary and this technique is compatible with a wider class of systems where focusing is not possible. While a few previous studies used interferometers~\cite{olbright1986interferometric,boudebs2001third,rodriguez2005new,dancus2013single}, they typically did not exploit images~\cite{olbright1986interferometric} or they relied on a pump-probe approach~\cite{boudebs2001third,rodriguez2005new,dancus2013single} (cross-Kerr effect) that would complicate the high-precision resonant nonlinear spectroscopy presented here. Moreover, we note that in parallel to our work an equivalent single-beam method has been developed, benchmarked and used on a single, non-Rydberg resonance in hot atomic vapors~\cite{aladjidi2022transit}. Importantly for our system, we significantly improved the phase resolution (e.g. $\times 2$ relative to Ref~\cite{boudebs2001third} and $\times 10$ relative to Ref~\cite{aladjidi2022transit}).


Nonlinear effects in excitonic spectra can be modeled in the framework of the so-called real density matrix approach (RDMA)~\cite{zielinska2019nonlinear}. The RDMA provides analytical expressions for the optical response of any semiconductor crystal using a small number of well-known parameters (e.g. effective masses, gap energy, dielectric constant). It can include Rydberg excitons of arbitrarily high principal quantum numbers, includes the case of indirect interband transitions, takes into account the effects of an anisotropic dispersion and the coherence of the electron and the hole with the radiation field. The total refraction index for an average intensity $I$ inside a crystal of the length L is given by $n^2=\epsilon_b+\chi^{(1)}+\chi^{(3)}(I)$, where $\chi^{(1)}$ and $\chi^{(3)}(I)$ are the linear and nonlinear parts of the susceptibility, allowing one to calculate the nonlinear phase shift $\Delta \phi = \frac{\omega L}{c}\left[n(I)-n(0)\right]$. The analytical expressions for the linear and nonlinear susceptibilities are derived in the Supplementary, where several strategies are discussed to take into account the Rydberg blockade effect. One approach is to treat exciton-exciton interactions as a broadening mechanism, from which the known Rydberg scaling laws~\cite{heckotter2017scaling} correctly lead to the prediction of an exciton density decreasing as $n^{-7}$ on resonance. A complementary approach consists in using a saturable function $f(I)=\frac{\alpha I}{1+I/I_{sat}}$ to scale either the constant $\chi^{(3)}_0$ or the oscillator strengths ${\mathcal F}_{nn'}$, mimicking the blockade. Both approaches show a good agreement with the measurement. However, the broadening method introduces some distortions to the line shape between each resonances that are not visible in the experiment. This is unlike with the saturable approach and we therefore favored it in the theoretical predictions shown here.

\begin{figure}[t!]
\begin{center}
  \includegraphics[width=1.0\linewidth]{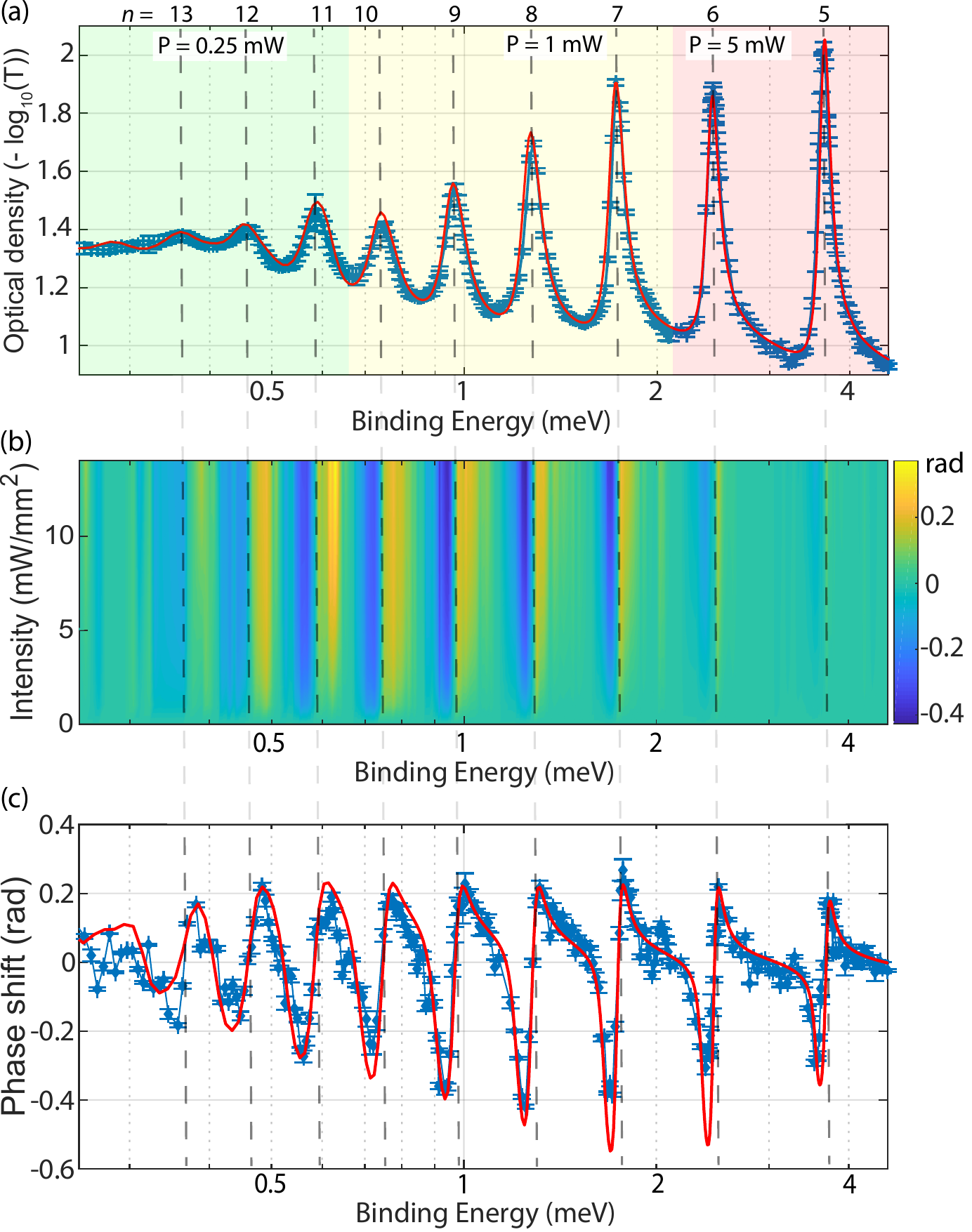}
  \caption{(a) Optical density spectrum of the yellow Rydberg series, plotted versus the exciton binding energy $E_b = E_{gap} - \hbar\omega$. Blue dots: experimental data (including error bars), red line: theory. Different powers were used for different states. (b) Color map of the nonlinear phase shift, plotted against the exciton binding energy and the intensity. It is zoomed on the low intensities, so that the low n-sates are far from saturation while the high n-states are rapidly saturated.  (c) Maximum (saturated signal) nonlinear phase shift. Blue dots: experimental data (including error bars), red line: theory.}
  \label{fig2}
\end{center}
\vspace{-5mm}
\end{figure}

Figure~\ref{fig2} (a) presents the absorption spectrum as a function of the exciton binding energy $E_b = E_{gap} - \hbar\omega$ ($E_{gap} = 2.1721~\si{\electronvolt}$), obtained by measuring the sample transmission at various energies $\hbar\omega$. Principal quantum numbers up to $n = 13$ are clearly observed. We attribute the lack of higher states to the relatively high temperature of $4~\si{\kelvin} \simeq 0.35~\si{\milli\electronvolt}$ that would ionize states $n > 14$. The fitting of our theoretical model indicates that the overall effect of temperature and sample quality is equal to a constant $21~\si{\micro\electronvolt}$ broadening of all excitonic lines. Due to the extreme $n$-dependence of the nonlinear absorption~\cite{kazimierczuk2014giant}, different laser powers were used for different states to avoid bleaching the high $n$ while keeping a sufficient signal across the large absorption of the low-$n$ peaks, as the color code in figure~\ref{fig2} (a) indicates. \ignore{The blockade-induced nonlinear absorption is therefore present and visible as a discontinuity in the peaks height when the power is changed.} As nonlinear dissipation has already been thoroughly studied in $\mathrm{Cu_2O}$, here we focus instead on the nonlinear dispersion. 

Figure~\ref{fig2} (b) is an example of observed self-Kerr phase shifts in the energy-intensity plane, zoomed on low laser intensities ($0-14~\si{\milli\watt/\milli\meter^2}$). As is visible by the increase of signal with $n$, higher $n$P states require less intensity to induce a phase shift\ignore{ (for example, compare the $n=5-6$ signal contrast to $n=9-10$)}. \ignore{Moreover, a saturation effect is also present that starts at a lower intensity $I_{sat}$ for higher values of $n$, as will be described later.} Additionally, the maximum phase shift $\Delta\phi_{max}$ spectrum is presented in figure~\ref{fig2} (c). It is maximum in the sense that we ensure saturation is reached: $\Delta\phi_{max} = \Delta\phi(I\gg I_{sat})-\Delta\phi(0)$ and the system cannot produce larger shifts regardless of the power used. We observe the typical positive-negative phase shift around each $n$P state, with a sharp zero-crossing exactly on resonance. The peak-to-peak amplitude first grows slowly with $n$ due to the increased nonlinear optical index $\mathrm{n_2}$. It then decreases to completely vanish at $n=14$. This is in good agreement with the observed broadening of the $n=11-13$ states and the absence of $n=14$ in the absorption. While $\Delta\phi_{max}$ has modest values of order $\pm 0.25$ rad, it is obtained at extremely low input intensities of order $1~\si{\milli\watt/\milli\meter^2}$ around $n = 10$. This is indicative of an extremely large nonlinearity near resonance, albeit one that saturates quickly.

\begin{figure}[t!]
\begin{center}
  \includegraphics[width=1.0\linewidth]{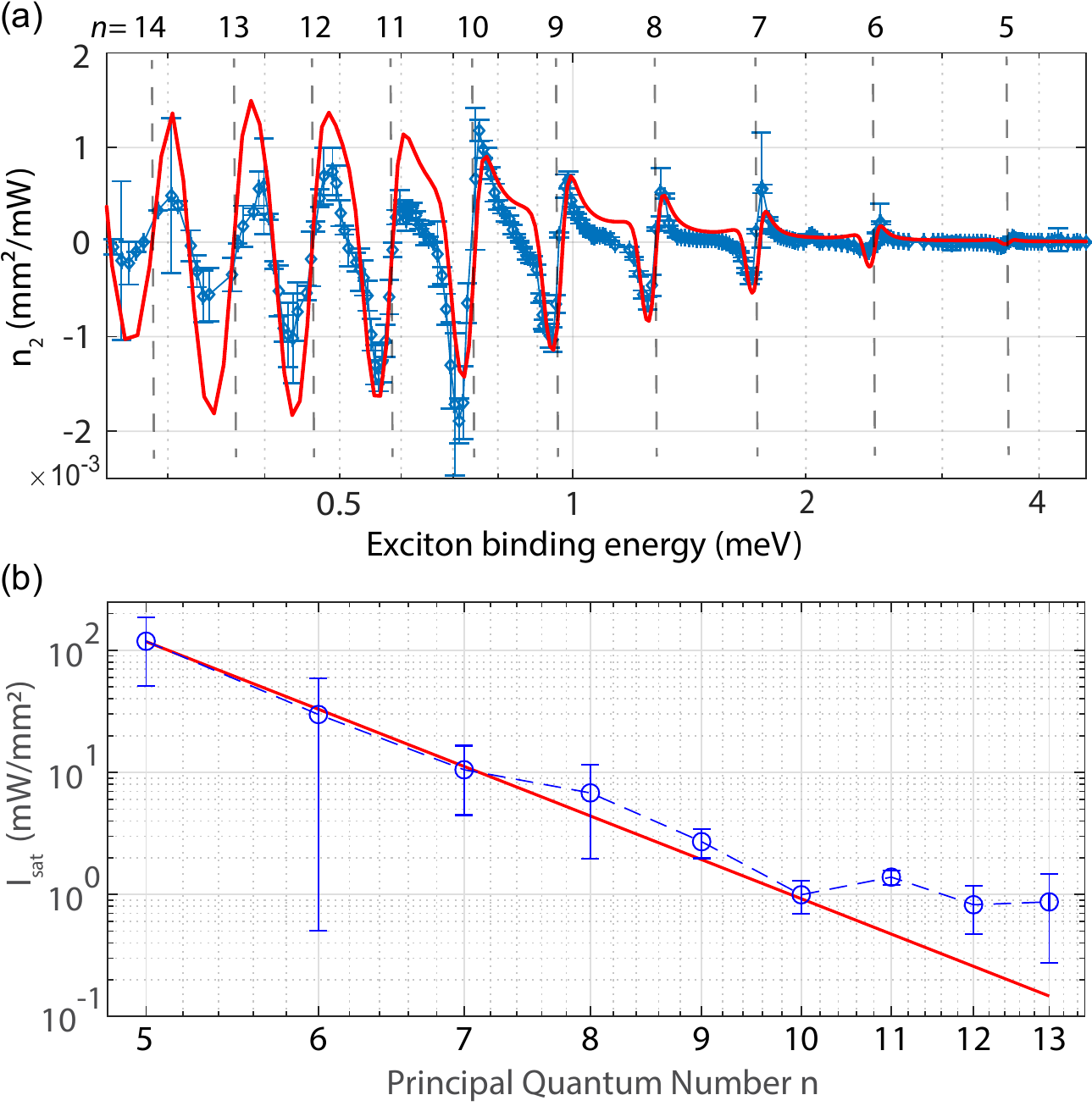}
  \caption{(a) Nonlinear index $\mathrm{n_2}$ extracted form the low-power derivative of $\Delta\phi(I)$. Blue dots: experimental data (including error bars), red line: theory. Note that here the state $n=14$ is visible. (b) A power law fit of $I_{sat}(n)$ indicates an exponent of $-6.9 \pm 0.2$. The red line is the $n^{-7}$ scaling law, almost identical to the best fit.}
  \label{fig3}
\end{center}
\vspace{-5mm}
\end{figure}

As saturation is present for all states beyond some intensity $I_{sat}$, we fit $\Delta\phi(I)$ with a saturable function $f(I) = \frac{\alpha I}{(1+I/I_{sat})}$ for each laser energy $\hbar\omega$. The nonlinear index $\mathrm{n_2}$ is calculated from the fit parameter $\alpha$ while the saturation intensity $I_{sat}$ is obtained directly.

\textit{Nonlinear index} - In a transparent medium far from saturation, the nonlinear phase shift is $\Delta\phi(I) = kLI\mathrm{n_2}$, where $k=\frac{2\pi}{\lambda}$ is the light wavevector. Therefore, $\mathrm{n_2} = \frac{1}{kL}\frac{\partial\Delta\phi}{\partial I}|_{I \ll I_{sat}} = \frac{\alpha}{kL}$. However, $\mathrm{Cu_2O}$ is not transparent and one has to take into account the reduced intensity inside the crystal. Rather than taking into account the full nonlinear intensity variation $I(z)$ along the propagation axis z, we make the approximation of a linear absorption $I(z)=I_0 e^{-z/z_0}$, where $z_0$ is the linear absorption length. This is justified as $\alpha=\frac{\partial\Delta\phi}{\partial I}|_{I_0 \to 0}$ is the derivative of $\Delta\phi(I)$ taken in the low intensity limit where nonlinear absorption is negligible. In that case we find $\mathrm{n_2} = \frac{\alpha}{kz_0}$, where $z_0 = -\frac{L}{\ln T} < L$ is inversely proportional to the optical density measured at low power. The resulting $\mathrm{n_2}(\hbar\omega)$ spectrum is shown in figure~\ref{fig3} (a). Interestingly, the peak-to-peak $\mathrm{n_2}$ amplitude rapidly increases with $n$ and reaches a maximum at $n=10$ of order $\mathrm{n_2 = }\si{10^{-3}\milli\meter^2/\milli\watt}$. This is about 2-4 orders of magnitude larger than in typical atomic systems~\cite{mccormick2004saturable,wang2020measurement,aladjidi2022transit} and 14 orders of magnitude beyond typical nonlinear crystals. For $n\geq 11$ the measured $\mathrm{n_2}$ falls off, likely due to the broadening from the finite crystal temperature and impurities. Note that here the $n=14$ signal is present, unlike in the optical density data: as extracting $\mathrm{n_2}$ only requires the low-intensity part of the pictures ($0\sim 0.5~\si{\milli\watt/\milli\meter^2}$ for $n=14$) we infer that, on top of being broadened by thermal phonons, $n=14$ is at least partly blockaded in figure~\ref{fig2} (a) and saturates too fast to show a significant signal in figure~\ref{fig2} (b-c).

\textit{Saturation} - The saturation intensity $I_{sat}(\omega)$ varies rapidly with the energy, going from $I_{sat}\rightarrow\infty$ far from resonance to a local minimum $I_{sat}^{min}$ around each resonance, as is typical of saturable systems. We focus here on the vicinity of $I_{sat}^{min}$ as its variation with $n$ brings interesting insights. We define $I_{sat}(n) = <I_{sat}(\omega)>_n$ where $<...>_n$ is the average over $\omega$ within a FWHM of the $n$P absorption peak. As visible in figure~\ref{fig3} (b), $I_{sat}(n)$ is of order $100~\si{\milli\watt/\milli\meter^{-2}}$ for $n=5$ but decreases rapidly, reaching $I_{sat} \lesssim 1~\si{\milli\watt/\milli\meter^{-2}}$ for $n\geq 10$. To find the origin of this rapid variation we fit a power law $A (n-\delta_\ell)^b$, where $\delta_{\ell=1} = 0.34$ is the quantum defect for P-states in $\mathrm{Cu_2O}$~\cite{heckotter2017scaling}. The exponent is found to be $b = -6.9 \pm 0.2$. This suggests the rapid saturation is a Rydberg blockade effect. Indeed, the total Kerr shift is proportional to density of excitons. However, due to the Rydberg blockade, the exciton density saturates and its maximum is the inverse blockade volume, $\rho_{max} = 1/V_{B} \propto n^{-7}$~\cite{kazimierczuk2014giant}. Therefore, Rydberg blockade implies that the Kerr shift saturates at some intensity $I_{sat} \propto \rho_{max} \propto n^{-7}$, as observed.

In conclusion, we measure the saturable nonlinear refraction induced by Rydberg excitons in $\mathrm{Cu_2O}$. Our approach is radically different from the traditional z-scan technique: by directly imaging the Kerr-induced phase shift $\Delta\phi(I)$, we simultaneously access the nonlinear index $\mathrm{n_2}$ and the saturation intensity $I_{sat}$. We reach the high precision of $\pm 0.01$ rad, allowing to resolve our relatively small signal. Interestingly, $\Delta\phi$ always remains small in spite of the gigantic $\mathrm{n_2}$ values reached at high principal quantum number because of a rapid saturation. We find this saturation is due to Rydberg blockade~\cite{kazimierczuk2014giant} inducing a saturation intensity $I_{sat} \propto n^{-7}$, which hinders large phase shifts. Incorporating this observed saturation into the  model developed in Ref.~\cite{zielinska2019nonlinear} yields excellent agreement between the theoretical predictions and the experimental results. To the best of our knowledge, electromagnetically-induced transparency (EIT) in cold atoms~\cite{hau1999light,fleischhauer2005electromagnetically,gorshkov2011photon,peyronel2012quantum} is the only scheme that has displayed a Kerr coefficient larger than the value of $\si{10^{-3}\milli\meter^2/\milli\watt}$ reported here.

Our results therefore underline that Rydberg excitons are a strong candidate for solid-state nonlinear quantum optics. While Rydberg blockade and absorption are a limitation in our simple one-photon study, the situation would be completely different with a resonant two-photon strategy in the fashion of Rydberg EIT~\cite{gorshkov2011photon,peyronel2012quantum}: the absorption would be suppressed while the blockade would massively enhance the Kerr coefficient, rather than hinder it. Therefore, EIT has the potential to reach the quantum regime of Kerr nonlinearity, unlike one-photon approaches. Recent theoretical studies pave the way toward such goals~\cite{walther2018giant,walther2021nonclassical}, using both intra-series~\cite{ziemkiewicz2020electromagnetically} and inter-series~\cite{kruger2019interseries,rommel2021interseries} Rydberg exciton transitions, including the non-trivial role of phonons~\cite{walther2020controlling}. With promising new results for intra-series coupling in the microwave domain~\cite{gallagher2022microwave} and recent advances in THz sources~\cite{curwen2019broadband,wu2021tunable} enabling to explore transitions in the $1 \sim 10~\si{\milli\electronvolt}$ range, experimentally probing two-photon strategies is likely to soon yield important results for solid-state Rydberg physics.\\



We thank Quentin Glorieux and Murad Abuzarli for useful discussions. This work has been supported by the ANR grant ANR-21-CE47-0008 (PIONEEReX).

\bibliographystyle{apsrev4-2}
\bibliography{Biblio}

\begin{thebibliography}{43}%
\makeatletter
\providecommand \@ifxundefined [1]{%
 \@ifx{#1\undefined}
}%
\providecommand \@ifnum [1]{%
 \ifnum #1\expandafter \@firstoftwo
 \else \expandafter \@secondoftwo
 \fi
}%
\providecommand \@ifx [1]{%
 \ifx #1\expandafter \@firstoftwo
 \else \expandafter \@secondoftwo
 \fi
}%
\providecommand \natexlab [1]{#1}%
\providecommand \enquote  [1]{``#1''}%
\providecommand \bibnamefont  [1]{#1}%
\providecommand \bibfnamefont [1]{#1}%
\providecommand \citenamefont [1]{#1}%
\providecommand \href@noop [0]{\@secondoftwo}%
\providecommand \href [0]{\begingroup \@sanitize@url \@href}%
\providecommand \@href[1]{\@@startlink{#1}\@@href}%
\providecommand \@@href[1]{\endgroup#1\@@endlink}%
\providecommand \@sanitize@url [0]{\catcode `\\12\catcode `\$12\catcode
  `\&12\catcode `\#12\catcode `\^12\catcode `\_12\catcode `\%12\relax}%
\providecommand \@@startlink[1]{}%
\providecommand \@@endlink[0]{}%
\providecommand \url  [0]{\begingroup\@sanitize@url \@url }%
\providecommand \@url [1]{\endgroup\@href {#1}{\urlprefix }}%
\providecommand \urlprefix  [0]{URL }%
\providecommand \Eprint [0]{\href }%
\providecommand \doibase [0]{https://doi.org/}%
\providecommand \selectlanguage [0]{\@gobble}%
\providecommand \bibinfo  [0]{\@secondoftwo}%
\providecommand \bibfield  [0]{\@secondoftwo}%
\providecommand \translation [1]{[#1]}%
\providecommand \BibitemOpen [0]{}%
\providecommand \bibitemStop [0]{}%
\providecommand \bibitemNoStop [0]{.\EOS\space}%
\providecommand \EOS [0]{\spacefactor3000\relax}%
\providecommand \BibitemShut  [1]{\csname bibitem#1\endcsname}%
\let\auto@bib@innerbib\@empty
\bibitem [{\citenamefont {Fleischhauer}\ \emph {et~al.}(2005)\citenamefont
  {Fleischhauer}, \citenamefont {Imamoglu},\ and\ \citenamefont
  {Marangos}}]{fleischhauer2005electromagnetically}%
  \BibitemOpen
  \bibfield  {author} {\bibinfo {author} {\bibfnamefont {M.}~\bibnamefont
  {Fleischhauer}}, \bibinfo {author} {\bibfnamefont {A.}~\bibnamefont
  {Imamoglu}},\ and\ \bibinfo {author} {\bibfnamefont {J.~P.}\ \bibnamefont
  {Marangos}},\ }\href@noop {} {\bibfield  {journal} {\bibinfo  {journal}
  {Reviews of modern physics}\ }\textbf {\bibinfo {volume} {77}},\ \bibinfo
  {pages} {633} (\bibinfo {year} {2005})}\BibitemShut {NoStop}%
\bibitem [{\citenamefont {Gorshkov}\ \emph {et~al.}(2011)\citenamefont
  {Gorshkov}, \citenamefont {Otterbach}, \citenamefont {Fleischhauer},
  \citenamefont {Pohl},\ and\ \citenamefont {Lukin}}]{gorshkov2011photon}%
  \BibitemOpen
  \bibfield  {author} {\bibinfo {author} {\bibfnamefont {A.~V.}\ \bibnamefont
  {Gorshkov}}, \bibinfo {author} {\bibfnamefont {J.}~\bibnamefont {Otterbach}},
  \bibinfo {author} {\bibfnamefont {M.}~\bibnamefont {Fleischhauer}}, \bibinfo
  {author} {\bibfnamefont {T.}~\bibnamefont {Pohl}},\ and\ \bibinfo {author}
  {\bibfnamefont {M.~D.}\ \bibnamefont {Lukin}},\ }\href@noop {} {\bibfield
  {journal} {\bibinfo  {journal} {Physical review letters}\ }\textbf {\bibinfo
  {volume} {107}},\ \bibinfo {pages} {133602} (\bibinfo {year}
  {2011})}\BibitemShut {NoStop}%
\bibitem [{\citenamefont {Peyronel}\ \emph {et~al.}(2012)\citenamefont
  {Peyronel}, \citenamefont {Firstenberg}, \citenamefont {Liang}, \citenamefont
  {Hofferberth}, \citenamefont {Gorshkov}, \citenamefont {Pohl}, \citenamefont
  {Lukin},\ and\ \citenamefont {Vuleti{\'c}}}]{peyronel2012quantum}%
  \BibitemOpen
  \bibfield  {author} {\bibinfo {author} {\bibfnamefont {T.}~\bibnamefont
  {Peyronel}}, \bibinfo {author} {\bibfnamefont {O.}~\bibnamefont
  {Firstenberg}}, \bibinfo {author} {\bibfnamefont {Q.-Y.}\ \bibnamefont
  {Liang}}, \bibinfo {author} {\bibfnamefont {S.}~\bibnamefont {Hofferberth}},
  \bibinfo {author} {\bibfnamefont {A.~V.}\ \bibnamefont {Gorshkov}}, \bibinfo
  {author} {\bibfnamefont {T.}~\bibnamefont {Pohl}}, \bibinfo {author}
  {\bibfnamefont {M.~D.}\ \bibnamefont {Lukin}},\ and\ \bibinfo {author}
  {\bibfnamefont {V.}~\bibnamefont {Vuleti{\'c}}},\ }\href@noop {} {\bibfield
  {journal} {\bibinfo  {journal} {Nature}\ }\textbf {\bibinfo {volume} {488}},\
  \bibinfo {pages} {57} (\bibinfo {year} {2012})}\BibitemShut {NoStop}%
\bibitem [{\citenamefont {A{\ss}mann}\ and\ \citenamefont
  {Bayer}(2020)}]{assmann2020semiconductor}%
  \BibitemOpen
  \bibfield  {author} {\bibinfo {author} {\bibfnamefont {M.}~\bibnamefont
  {A{\ss}mann}}\ and\ \bibinfo {author} {\bibfnamefont {M.}~\bibnamefont
  {Bayer}},\ }\href@noop {} {\bibfield  {journal} {\bibinfo  {journal}
  {Advanced Quantum Technologies}\ }\textbf {\bibinfo {volume} {3}},\ \bibinfo
  {pages} {1900134} (\bibinfo {year} {2020})}\BibitemShut {NoStop}%
\bibitem [{\citenamefont {Taylor}\ \emph {et~al.}(2021)\citenamefont {Taylor},
  \citenamefont {Goswami}, \citenamefont {Walther}, \citenamefont {Spanner},
  \citenamefont {Simon},\ and\ \citenamefont {Heshami}}]{taylor2021simulation}%
  \BibitemOpen
  \bibfield  {author} {\bibinfo {author} {\bibfnamefont {J.}~\bibnamefont
  {Taylor}}, \bibinfo {author} {\bibfnamefont {S.}~\bibnamefont {Goswami}},
  \bibinfo {author} {\bibfnamefont {V.}~\bibnamefont {Walther}}, \bibinfo
  {author} {\bibfnamefont {M.}~\bibnamefont {Spanner}}, \bibinfo {author}
  {\bibfnamefont {C.}~\bibnamefont {Simon}},\ and\ \bibinfo {author}
  {\bibfnamefont {K.}~\bibnamefont {Heshami}},\ }\href@noop {} {\bibfield
  {journal} {\bibinfo  {journal} {arXiv preprint arXiv:2107.02273}\ } (\bibinfo
  {year} {2021})}\BibitemShut {NoStop}%
\bibitem [{\citenamefont {Walther}\ \emph {et~al.}(2018)\citenamefont
  {Walther}, \citenamefont {Johne},\ and\ \citenamefont
  {Pohl}}]{walther2018giant}%
  \BibitemOpen
  \bibfield  {author} {\bibinfo {author} {\bibfnamefont {V.}~\bibnamefont
  {Walther}}, \bibinfo {author} {\bibfnamefont {R.}~\bibnamefont {Johne}},\
  and\ \bibinfo {author} {\bibfnamefont {T.}~\bibnamefont {Pohl}},\ }\href@noop
  {} {\bibfield  {journal} {\bibinfo  {journal} {Nature communications}\
  }\textbf {\bibinfo {volume} {9}},\ \bibinfo {pages} {1} (\bibinfo {year}
  {2018})}\BibitemShut {NoStop}%
\bibitem [{\citenamefont {Walther}\ \emph {et~al.}(2021)\citenamefont
  {Walther}, \citenamefont {Zhang}, \citenamefont {Yelin},\ and\ \citenamefont
  {Pohl}}]{walther2021nonclassical}%
  \BibitemOpen
  \bibfield  {author} {\bibinfo {author} {\bibfnamefont {V.}~\bibnamefont
  {Walther}}, \bibinfo {author} {\bibfnamefont {L.}~\bibnamefont {Zhang}},
  \bibinfo {author} {\bibfnamefont {S.~F.}\ \bibnamefont {Yelin}},\ and\
  \bibinfo {author} {\bibfnamefont {T.}~\bibnamefont {Pohl}},\ }\href@noop {}
  {\bibfield  {journal} {\bibinfo  {journal} {arXiv preprint arXiv:2102.10350}\
  } (\bibinfo {year} {2021})}\BibitemShut {NoStop}%
\bibitem [{\citenamefont {Kazimierczuk}\ \emph {et~al.}(2014)\citenamefont
  {Kazimierczuk}, \citenamefont {Fr{\"o}hlich}, \citenamefont {Scheel},
  \citenamefont {Stolz},\ and\ \citenamefont {Bayer}}]{kazimierczuk2014giant}%
  \BibitemOpen
  \bibfield  {author} {\bibinfo {author} {\bibfnamefont {T.}~\bibnamefont
  {Kazimierczuk}}, \bibinfo {author} {\bibfnamefont {D.}~\bibnamefont
  {Fr{\"o}hlich}}, \bibinfo {author} {\bibfnamefont {S.}~\bibnamefont
  {Scheel}}, \bibinfo {author} {\bibfnamefont {H.}~\bibnamefont {Stolz}},\ and\
  \bibinfo {author} {\bibfnamefont {M.}~\bibnamefont {Bayer}},\ }\href@noop {}
  {\bibfield  {journal} {\bibinfo  {journal} {Nature}\ }\textbf {\bibinfo
  {volume} {514}},\ \bibinfo {pages} {343} (\bibinfo {year}
  {2014})}\BibitemShut {NoStop}%
\bibitem [{\citenamefont {Heck{\"o}tter}\ \emph {et~al.}(2020)\citenamefont
  {Heck{\"o}tter}, \citenamefont {Janas}, \citenamefont {Schwartz},
  \citenamefont {A{\ss}mann},\ and\ \citenamefont
  {Bayer}}]{heckotter2020experimental}%
  \BibitemOpen
  \bibfield  {author} {\bibinfo {author} {\bibfnamefont {J.}~\bibnamefont
  {Heck{\"o}tter}}, \bibinfo {author} {\bibfnamefont {D.}~\bibnamefont
  {Janas}}, \bibinfo {author} {\bibfnamefont {R.}~\bibnamefont {Schwartz}},
  \bibinfo {author} {\bibfnamefont {M.}~\bibnamefont {A{\ss}mann}},\ and\
  \bibinfo {author} {\bibfnamefont {M.}~\bibnamefont {Bayer}},\ }\href@noop {}
  {\bibfield  {journal} {\bibinfo  {journal} {Physical Review B}\ }\textbf
  {\bibinfo {volume} {101}},\ \bibinfo {pages} {235207} (\bibinfo {year}
  {2020})}\BibitemShut {NoStop}%
\bibitem [{\citenamefont {Versteegh}\ \emph {et~al.}(2021)\citenamefont
  {Versteegh}, \citenamefont {Steinhauer}, \citenamefont {Bajo}, \citenamefont
  {Lettner}, \citenamefont {Soro}, \citenamefont {Romanova}, \citenamefont
  {Gyger}, \citenamefont {Schweickert}, \citenamefont {Mysyrowicz},\ and\
  \citenamefont {Zwiller}}]{versteegh2021giant}%
  \BibitemOpen
  \bibfield  {author} {\bibinfo {author} {\bibfnamefont {M.~A.}\ \bibnamefont
  {Versteegh}}, \bibinfo {author} {\bibfnamefont {S.}~\bibnamefont
  {Steinhauer}}, \bibinfo {author} {\bibfnamefont {J.}~\bibnamefont {Bajo}},
  \bibinfo {author} {\bibfnamefont {T.}~\bibnamefont {Lettner}}, \bibinfo
  {author} {\bibfnamefont {A.}~\bibnamefont {Soro}}, \bibinfo {author}
  {\bibfnamefont {A.}~\bibnamefont {Romanova}}, \bibinfo {author}
  {\bibfnamefont {S.}~\bibnamefont {Gyger}}, \bibinfo {author} {\bibfnamefont
  {L.}~\bibnamefont {Schweickert}}, \bibinfo {author} {\bibfnamefont
  {A.}~\bibnamefont {Mysyrowicz}},\ and\ \bibinfo {author} {\bibfnamefont
  {V.}~\bibnamefont {Zwiller}},\ }\href@noop {} {\bibfield  {journal} {\bibinfo
   {journal} {Physical Review B}\ }\textbf {\bibinfo {volume} {104}},\ \bibinfo
  {pages} {245206} (\bibinfo {year} {2021})}\BibitemShut {NoStop}%
\bibitem [{\citenamefont {Heck{\"o}tter}\ \emph {et~al.}(2017)\citenamefont
  {Heck{\"o}tter}, \citenamefont {Freitag}, \citenamefont {Fr{\"o}hlich},
  \citenamefont {A{\ss}mann}, \citenamefont {Bayer}, \citenamefont {Semina},\
  and\ \citenamefont {Glazov}}]{heckotter2017scaling}%
  \BibitemOpen
  \bibfield  {author} {\bibinfo {author} {\bibfnamefont {J.}~\bibnamefont
  {Heck{\"o}tter}}, \bibinfo {author} {\bibfnamefont {M.}~\bibnamefont
  {Freitag}}, \bibinfo {author} {\bibfnamefont {D.}~\bibnamefont
  {Fr{\"o}hlich}}, \bibinfo {author} {\bibfnamefont {M.}~\bibnamefont
  {A{\ss}mann}}, \bibinfo {author} {\bibfnamefont {M.}~\bibnamefont {Bayer}},
  \bibinfo {author} {\bibfnamefont {M.}~\bibnamefont {Semina}},\ and\ \bibinfo
  {author} {\bibfnamefont {M.}~\bibnamefont {Glazov}},\ }\href@noop {}
  {\bibfield  {journal} {\bibinfo  {journal} {Physical Review B}\ }\textbf
  {\bibinfo {volume} {96}},\ \bibinfo {pages} {125142} (\bibinfo {year}
  {2017})}\BibitemShut {NoStop}%
\bibitem [{\citenamefont {Gr{\"u}nwald}\ \emph {et~al.}(2016)\citenamefont
  {Gr{\"u}nwald}, \citenamefont {A{\ss}mann}, \citenamefont {Heck{\"o}tter},
  \citenamefont {Fr{\"o}hlich}, \citenamefont {Bayer}, \citenamefont {Stolz},\
  and\ \citenamefont {Scheel}}]{grunwald2016signatures}%
  \BibitemOpen
  \bibfield  {author} {\bibinfo {author} {\bibfnamefont {P.}~\bibnamefont
  {Gr{\"u}nwald}}, \bibinfo {author} {\bibfnamefont {M.}~\bibnamefont
  {A{\ss}mann}}, \bibinfo {author} {\bibfnamefont {J.}~\bibnamefont
  {Heck{\"o}tter}}, \bibinfo {author} {\bibfnamefont {D.}~\bibnamefont
  {Fr{\"o}hlich}}, \bibinfo {author} {\bibfnamefont {M.}~\bibnamefont {Bayer}},
  \bibinfo {author} {\bibfnamefont {H.}~\bibnamefont {Stolz}},\ and\ \bibinfo
  {author} {\bibfnamefont {S.}~\bibnamefont {Scheel}},\ }\href@noop {}
  {\bibfield  {journal} {\bibinfo  {journal} {Physical review letters}\
  }\textbf {\bibinfo {volume} {117}},\ \bibinfo {pages} {133003} (\bibinfo
  {year} {2016})}\BibitemShut {NoStop}%
\bibitem [{\citenamefont {Zieli{\'n}ska-Raczy{\'n}ska}\ \emph
  {et~al.}(2019)\citenamefont {Zieli{\'n}ska-Raczy{\'n}ska}, \citenamefont
  {Czajkowski}, \citenamefont {Karpi{\'n}ski},\ and\ \citenamefont
  {Ziemkiewicz}}]{zielinska2019nonlinear}%
  \BibitemOpen
  \bibfield  {author} {\bibinfo {author} {\bibfnamefont {S.}~\bibnamefont
  {Zieli{\'n}ska-Raczy{\'n}ska}}, \bibinfo {author} {\bibfnamefont
  {G.}~\bibnamefont {Czajkowski}}, \bibinfo {author} {\bibfnamefont
  {K.}~\bibnamefont {Karpi{\'n}ski}},\ and\ \bibinfo {author} {\bibfnamefont
  {D.}~\bibnamefont {Ziemkiewicz}},\ }\href@noop {} {\bibfield  {journal}
  {\bibinfo  {journal} {Physical Review B}\ }\textbf {\bibinfo {volume} {99}},\
  \bibinfo {pages} {245206} (\bibinfo {year} {2019})}\BibitemShut {NoStop}%
\bibitem [{\citenamefont {Walther}\ \emph {et~al.}(2020)\citenamefont
  {Walther}, \citenamefont {Gr{\"u}nwald},\ and\ \citenamefont
  {Pohl}}]{walther2020controlling}%
  \BibitemOpen
  \bibfield  {author} {\bibinfo {author} {\bibfnamefont {V.}~\bibnamefont
  {Walther}}, \bibinfo {author} {\bibfnamefont {P.}~\bibnamefont
  {Gr{\"u}nwald}},\ and\ \bibinfo {author} {\bibfnamefont {T.}~\bibnamefont
  {Pohl}},\ }\href@noop {} {\bibfield  {journal} {\bibinfo  {journal} {Physical
  Review Letters}\ }\textbf {\bibinfo {volume} {125}},\ \bibinfo {pages}
  {173601} (\bibinfo {year} {2020})}\BibitemShut {NoStop}%
\bibitem [{\citenamefont {Walther}\ and\ \citenamefont
  {Pohl}(2020)}]{walther2020plasma}%
  \BibitemOpen
  \bibfield  {author} {\bibinfo {author} {\bibfnamefont {V.}~\bibnamefont
  {Walther}}\ and\ \bibinfo {author} {\bibfnamefont {T.}~\bibnamefont {Pohl}},\
  }\href@noop {} {\bibfield  {journal} {\bibinfo  {journal} {Physical Review
  Letters}\ }\textbf {\bibinfo {volume} {125}},\ \bibinfo {pages} {097401}
  (\bibinfo {year} {2020})}\BibitemShut {NoStop}%
\bibitem [{\citenamefont {Khazali}\ \emph {et~al.}(2017)\citenamefont
  {Khazali}, \citenamefont {Heshami},\ and\ \citenamefont
  {Simon}}]{khazali2017single}%
  \BibitemOpen
  \bibfield  {author} {\bibinfo {author} {\bibfnamefont {M.}~\bibnamefont
  {Khazali}}, \bibinfo {author} {\bibfnamefont {K.}~\bibnamefont {Heshami}},\
  and\ \bibinfo {author} {\bibfnamefont {C.}~\bibnamefont {Simon}},\
  }\href@noop {} {\bibfield  {journal} {\bibinfo  {journal} {Journal of Physics
  B: Atomic, Molecular and Optical Physics}\ }\textbf {\bibinfo {volume}
  {50}},\ \bibinfo {pages} {215301} (\bibinfo {year} {2017})}\BibitemShut
  {NoStop}%
\bibitem [{\citenamefont
  {Ziemkiewicz}(2020)}]{ziemkiewicz2020electromagnetically}%
  \BibitemOpen
  \bibfield  {author} {\bibinfo {author} {\bibfnamefont {D.}~\bibnamefont
  {Ziemkiewicz}},\ }\href@noop {} {\bibfield  {journal} {\bibinfo  {journal}
  {Entropy}\ }\textbf {\bibinfo {volume} {22}},\ \bibinfo {pages} {177}
  (\bibinfo {year} {2020})}\BibitemShut {NoStop}%
\bibitem [{\citenamefont {Farenbruch}\ \emph {et~al.}(2020)\citenamefont
  {Farenbruch}, \citenamefont {Fr{\"o}hlich}, \citenamefont {Yakovlev},\ and\
  \citenamefont {Bayer}}]{farenbruch2020rydberg}%
  \BibitemOpen
  \bibfield  {author} {\bibinfo {author} {\bibfnamefont {A.}~\bibnamefont
  {Farenbruch}}, \bibinfo {author} {\bibfnamefont {D.}~\bibnamefont
  {Fr{\"o}hlich}}, \bibinfo {author} {\bibfnamefont {D.~R.}\ \bibnamefont
  {Yakovlev}},\ and\ \bibinfo {author} {\bibfnamefont {M.}~\bibnamefont
  {Bayer}},\ }\href@noop {} {\bibfield  {journal} {\bibinfo  {journal}
  {Physical Review Letters}\ }\textbf {\bibinfo {volume} {125}},\ \bibinfo
  {pages} {207402} (\bibinfo {year} {2020})}\BibitemShut {NoStop}%
\bibitem [{\citenamefont {Rogers}\ \emph {et~al.}(2021)\citenamefont {Rogers},
  \citenamefont {Gallagher}, \citenamefont {Pizzey}, \citenamefont {Pritchett},
  \citenamefont {Adams}, \citenamefont {Jones}, \citenamefont {Hodges},
  \citenamefont {Langbein},\ and\ \citenamefont {Lynch}}]{rogers2021high}%
  \BibitemOpen
  \bibfield  {author} {\bibinfo {author} {\bibfnamefont {J.~P.}\ \bibnamefont
  {Rogers}}, \bibinfo {author} {\bibfnamefont {L.~A.}\ \bibnamefont
  {Gallagher}}, \bibinfo {author} {\bibfnamefont {D.}~\bibnamefont {Pizzey}},
  \bibinfo {author} {\bibfnamefont {J.~D.}\ \bibnamefont {Pritchett}}, \bibinfo
  {author} {\bibfnamefont {C.~S.}\ \bibnamefont {Adams}}, \bibinfo {author}
  {\bibfnamefont {M.}~\bibnamefont {Jones}}, \bibinfo {author} {\bibfnamefont
  {C.}~\bibnamefont {Hodges}}, \bibinfo {author} {\bibfnamefont
  {W.}~\bibnamefont {Langbein}},\ and\ \bibinfo {author} {\bibfnamefont
  {S.~A.}\ \bibnamefont {Lynch}},\ }\href@noop {} {\bibfield  {journal}
  {\bibinfo  {journal} {arXiv preprint arXiv:2111.13062}\ } (\bibinfo {year}
  {2021})}\BibitemShut {NoStop}%
\bibitem [{\citenamefont {Mund}\ \emph {et~al.}(2018)\citenamefont {Mund},
  \citenamefont {Fr{\"o}hlich}, \citenamefont {Yakovlev},\ and\ \citenamefont
  {Bayer}}]{mund2018high}%
  \BibitemOpen
  \bibfield  {author} {\bibinfo {author} {\bibfnamefont {J.}~\bibnamefont
  {Mund}}, \bibinfo {author} {\bibfnamefont {D.}~\bibnamefont {Fr{\"o}hlich}},
  \bibinfo {author} {\bibfnamefont {D.~R.}\ \bibnamefont {Yakovlev}},\ and\
  \bibinfo {author} {\bibfnamefont {M.}~\bibnamefont {Bayer}},\ }\href@noop {}
  {\bibfield  {journal} {\bibinfo  {journal} {Physical Review B}\ }\textbf
  {\bibinfo {volume} {98}},\ \bibinfo {pages} {085203} (\bibinfo {year}
  {2018})}\BibitemShut {NoStop}%
\bibitem [{\citenamefont {Farenbruch}\ \emph {et~al.}(2021)\citenamefont
  {Farenbruch}, \citenamefont {Fr{\"o}hlich}, \citenamefont {Stolz},
  \citenamefont {Yakovlev},\ and\ \citenamefont
  {Bayer}}]{farenbruch2021second}%
  \BibitemOpen
  \bibfield  {author} {\bibinfo {author} {\bibfnamefont {A.}~\bibnamefont
  {Farenbruch}}, \bibinfo {author} {\bibfnamefont {D.}~\bibnamefont
  {Fr{\"o}hlich}}, \bibinfo {author} {\bibfnamefont {H.}~\bibnamefont {Stolz}},
  \bibinfo {author} {\bibfnamefont {D.}~\bibnamefont {Yakovlev}},\ and\
  \bibinfo {author} {\bibfnamefont {M.}~\bibnamefont {Bayer}},\ }\href@noop {}
  {\bibfield  {journal} {\bibinfo  {journal} {Physical Review B}\ }\textbf
  {\bibinfo {volume} {104}},\ \bibinfo {pages} {075203} (\bibinfo {year}
  {2021})}\BibitemShut {NoStop}%
\bibitem [{\citenamefont {Heck{\"o}tter}\ \emph {et~al.}(2021)\citenamefont
  {Heck{\"o}tter}, \citenamefont {Walther}, \citenamefont {Scheel},
  \citenamefont {Bayer}, \citenamefont {Pohl},\ and\ \citenamefont
  {A{\ss}mann}}]{heckotter2021asymmetric}%
  \BibitemOpen
  \bibfield  {author} {\bibinfo {author} {\bibfnamefont {J.}~\bibnamefont
  {Heck{\"o}tter}}, \bibinfo {author} {\bibfnamefont {V.}~\bibnamefont
  {Walther}}, \bibinfo {author} {\bibfnamefont {S.}~\bibnamefont {Scheel}},
  \bibinfo {author} {\bibfnamefont {M.}~\bibnamefont {Bayer}}, \bibinfo
  {author} {\bibfnamefont {T.}~\bibnamefont {Pohl}},\ and\ \bibinfo {author}
  {\bibfnamefont {M.}~\bibnamefont {A{\ss}mann}},\ }\href@noop {} {\bibfield
  {journal} {\bibinfo  {journal} {Nature communications}\ }\textbf {\bibinfo
  {volume} {12}},\ \bibinfo {pages} {1} (\bibinfo {year} {2021})}\BibitemShut
  {NoStop}%
\bibitem [{\citenamefont {Gallagher}\ \emph {et~al.}(2022)\citenamefont
  {Gallagher}, \citenamefont {Rogers}, \citenamefont {Pritchett}, \citenamefont
  {Mistry}, \citenamefont {Pizzey}, \citenamefont {Adams}, \citenamefont
  {Jones}, \citenamefont {Gr{\"u}nwald}, \citenamefont {Walther}, \citenamefont
  {Hodges} \emph {et~al.}}]{gallagher2022microwave}%
  \BibitemOpen
  \bibfield  {author} {\bibinfo {author} {\bibfnamefont {L.~A.}\ \bibnamefont
  {Gallagher}}, \bibinfo {author} {\bibfnamefont {J.~P.}\ \bibnamefont
  {Rogers}}, \bibinfo {author} {\bibfnamefont {J.~D.}\ \bibnamefont
  {Pritchett}}, \bibinfo {author} {\bibfnamefont {R.~A.}\ \bibnamefont
  {Mistry}}, \bibinfo {author} {\bibfnamefont {D.}~\bibnamefont {Pizzey}},
  \bibinfo {author} {\bibfnamefont {C.~S.}\ \bibnamefont {Adams}}, \bibinfo
  {author} {\bibfnamefont {M.~P.}\ \bibnamefont {Jones}}, \bibinfo {author}
  {\bibfnamefont {P.}~\bibnamefont {Gr{\"u}nwald}}, \bibinfo {author}
  {\bibfnamefont {V.}~\bibnamefont {Walther}}, \bibinfo {author} {\bibfnamefont
  {C.}~\bibnamefont {Hodges}}, \emph {et~al.},\ }\href@noop {} {\bibfield
  {journal} {\bibinfo  {journal} {Physical Review Research}\ }\textbf {\bibinfo
  {volume} {4}},\ \bibinfo {pages} {013031} (\bibinfo {year}
  {2022})}\BibitemShut {NoStop}%
\bibitem [{\citenamefont {Brabec}\ \emph {et~al.}(1992)\citenamefont {Brabec},
  \citenamefont {Spielmann}, \citenamefont {Curley},\ and\ \citenamefont
  {Krausz}}]{brabec1992kerr}%
  \BibitemOpen
  \bibfield  {author} {\bibinfo {author} {\bibfnamefont {T.}~\bibnamefont
  {Brabec}}, \bibinfo {author} {\bibfnamefont {C.}~\bibnamefont {Spielmann}},
  \bibinfo {author} {\bibfnamefont {P.}~\bibnamefont {Curley}},\ and\ \bibinfo
  {author} {\bibfnamefont {F.}~\bibnamefont {Krausz}},\ }\href@noop {}
  {\bibfield  {journal} {\bibinfo  {journal} {Optics letters}\ }\textbf
  {\bibinfo {volume} {17}},\ \bibinfo {pages} {1292} (\bibinfo {year}
  {1992})}\BibitemShut {NoStop}%
\bibitem [{\citenamefont {Carusotto}\ and\ \citenamefont
  {Ciuti}(2013)}]{carusotto2013quantum}%
  \BibitemOpen
  \bibfield  {author} {\bibinfo {author} {\bibfnamefont {I.}~\bibnamefont
  {Carusotto}}\ and\ \bibinfo {author} {\bibfnamefont {C.}~\bibnamefont
  {Ciuti}},\ }\href@noop {} {\bibfield  {journal} {\bibinfo  {journal} {Reviews
  of Modern Physics}\ }\textbf {\bibinfo {volume} {85}},\ \bibinfo {pages}
  {299} (\bibinfo {year} {2013})}\BibitemShut {NoStop}%
\bibitem [{\citenamefont {Fontaine}\ \emph {et~al.}(2018)\citenamefont
  {Fontaine}, \citenamefont {Bienaim{\'e}}, \citenamefont {Pigeon},
  \citenamefont {Giacobino}, \citenamefont {Bramati},\ and\ \citenamefont
  {Glorieux}}]{fontaine2018observation}%
  \BibitemOpen
  \bibfield  {author} {\bibinfo {author} {\bibfnamefont {Q.}~\bibnamefont
  {Fontaine}}, \bibinfo {author} {\bibfnamefont {T.}~\bibnamefont
  {Bienaim{\'e}}}, \bibinfo {author} {\bibfnamefont {S.}~\bibnamefont
  {Pigeon}}, \bibinfo {author} {\bibfnamefont {E.}~\bibnamefont {Giacobino}},
  \bibinfo {author} {\bibfnamefont {A.}~\bibnamefont {Bramati}},\ and\ \bibinfo
  {author} {\bibfnamefont {Q.}~\bibnamefont {Glorieux}},\ }\href@noop {}
  {\bibfield  {journal} {\bibinfo  {journal} {Physical review letters}\
  }\textbf {\bibinfo {volume} {121}},\ \bibinfo {pages} {183604} (\bibinfo
  {year} {2018})}\BibitemShut {NoStop}%
\bibitem [{\citenamefont {Boulier}\ \emph {et~al.}(2020)\citenamefont
  {Boulier}, \citenamefont {Jacquet}, \citenamefont {Ma{\^\i}tre},
  \citenamefont {Lerario}, \citenamefont {Claude}, \citenamefont {Pigeon},
  \citenamefont {Glorieux}, \citenamefont {Amo}, \citenamefont {Bloch},
  \citenamefont {Bramati} \emph {et~al.}}]{boulier2020microcavity}%
  \BibitemOpen
  \bibfield  {author} {\bibinfo {author} {\bibfnamefont {T.}~\bibnamefont
  {Boulier}}, \bibinfo {author} {\bibfnamefont {M.~J.}\ \bibnamefont
  {Jacquet}}, \bibinfo {author} {\bibfnamefont {A.}~\bibnamefont
  {Ma{\^\i}tre}}, \bibinfo {author} {\bibfnamefont {G.}~\bibnamefont
  {Lerario}}, \bibinfo {author} {\bibfnamefont {F.}~\bibnamefont {Claude}},
  \bibinfo {author} {\bibfnamefont {S.}~\bibnamefont {Pigeon}}, \bibinfo
  {author} {\bibfnamefont {Q.}~\bibnamefont {Glorieux}}, \bibinfo {author}
  {\bibfnamefont {A.}~\bibnamefont {Amo}}, \bibinfo {author} {\bibfnamefont
  {J.}~\bibnamefont {Bloch}}, \bibinfo {author} {\bibfnamefont
  {A.}~\bibnamefont {Bramati}}, \emph {et~al.},\ }\href@noop {} {\bibfield
  {journal} {\bibinfo  {journal} {Advanced Quantum Technologies}\ }\textbf
  {\bibinfo {volume} {3}},\ \bibinfo {pages} {2000052} (\bibinfo {year}
  {2020})}\BibitemShut {NoStop}%
\bibitem [{\citenamefont {Chang}\ \emph {et~al.}(2014)\citenamefont {Chang},
  \citenamefont {Vuleti{\'c}},\ and\ \citenamefont {Lukin}}]{chang2014quantum}%
  \BibitemOpen
  \bibfield  {author} {\bibinfo {author} {\bibfnamefont {D.~E.}\ \bibnamefont
  {Chang}}, \bibinfo {author} {\bibfnamefont {V.}~\bibnamefont {Vuleti{\'c}}},\
  and\ \bibinfo {author} {\bibfnamefont {M.~D.}\ \bibnamefont {Lukin}},\
  }\href@noop {} {\bibfield  {journal} {\bibinfo  {journal} {Nature Photonics}\
  }\textbf {\bibinfo {volume} {8}},\ \bibinfo {pages} {685} (\bibinfo {year}
  {2014})}\BibitemShut {NoStop}%
\bibitem [{\citenamefont {Olbright}\ and\ \citenamefont
  {Peyghambarian}(1986)}]{olbright1986interferometric}%
  \BibitemOpen
  \bibfield  {author} {\bibinfo {author} {\bibfnamefont {G.}~\bibnamefont
  {Olbright}}\ and\ \bibinfo {author} {\bibfnamefont {N.}~\bibnamefont
  {Peyghambarian}},\ }\href@noop {} {\bibfield  {journal} {\bibinfo  {journal}
  {Applied physics letters}\ }\textbf {\bibinfo {volume} {48}},\ \bibinfo
  {pages} {1184} (\bibinfo {year} {1986})}\BibitemShut {NoStop}%
\bibitem [{\citenamefont {Sheik-Bahae}\ \emph {et~al.}(1989)\citenamefont
  {Sheik-Bahae}, \citenamefont {Said},\ and\ \citenamefont
  {Van~Stryland}}]{sheik1989high}%
  \BibitemOpen
  \bibfield  {author} {\bibinfo {author} {\bibfnamefont {M.}~\bibnamefont
  {Sheik-Bahae}}, \bibinfo {author} {\bibfnamefont {A.~A.}\ \bibnamefont
  {Said}},\ and\ \bibinfo {author} {\bibfnamefont {E.~W.}\ \bibnamefont
  {Van~Stryland}},\ }\href@noop {} {\bibfield  {journal} {\bibinfo  {journal}
  {Optics letters}\ }\textbf {\bibinfo {volume} {14}},\ \bibinfo {pages} {955}
  (\bibinfo {year} {1989})}\BibitemShut {NoStop}%
\bibitem [{\citenamefont {Boudebs}\ \emph {et~al.}(2001)\citenamefont
  {Boudebs}, \citenamefont {Chis},\ and\ \citenamefont
  {Phu}}]{boudebs2001third}%
  \BibitemOpen
  \bibfield  {author} {\bibinfo {author} {\bibfnamefont {G.}~\bibnamefont
  {Boudebs}}, \bibinfo {author} {\bibfnamefont {M.}~\bibnamefont {Chis}},\ and\
  \bibinfo {author} {\bibfnamefont {X.~N.}\ \bibnamefont {Phu}},\ }\href@noop
  {} {\bibfield  {journal} {\bibinfo  {journal} {JOSA B}\ }\textbf {\bibinfo
  {volume} {18}},\ \bibinfo {pages} {623} (\bibinfo {year} {2001})}\BibitemShut
  {NoStop}%
\bibitem [{\citenamefont {Rodr{\'\i}guez}\ \emph {et~al.}(2005)\citenamefont
  {Rodr{\'\i}guez}, \citenamefont {Simos}, \citenamefont {Sylla}, \citenamefont
  {Phu} \emph {et~al.}}]{rodriguez2005new}%
  \BibitemOpen
  \bibfield  {author} {\bibinfo {author} {\bibfnamefont {L.}~\bibnamefont
  {Rodr{\'\i}guez}}, \bibinfo {author} {\bibfnamefont {C.}~\bibnamefont
  {Simos}}, \bibinfo {author} {\bibfnamefont {M.}~\bibnamefont {Sylla}},
  \bibinfo {author} {\bibfnamefont {X.~N.}\ \bibnamefont {Phu}}, \emph
  {et~al.},\ }\href@noop {} {\bibfield  {journal} {\bibinfo  {journal} {Optics
  communications}\ }\textbf {\bibinfo {volume} {247}},\ \bibinfo {pages} {453}
  (\bibinfo {year} {2005})}\BibitemShut {NoStop}%
\bibitem [{\citenamefont {Dancus}\ \emph {et~al.}(2013)\citenamefont {Dancus},
  \citenamefont {Popescu},\ and\ \citenamefont {Petris}}]{dancus2013single}%
  \BibitemOpen
  \bibfield  {author} {\bibinfo {author} {\bibfnamefont {I.}~\bibnamefont
  {Dancus}}, \bibinfo {author} {\bibfnamefont {S.~T.}\ \bibnamefont
  {Popescu}},\ and\ \bibinfo {author} {\bibfnamefont {A.}~\bibnamefont
  {Petris}},\ }\href@noop {} {\bibfield  {journal} {\bibinfo  {journal} {Optics
  Express}\ }\textbf {\bibinfo {volume} {21}},\ \bibinfo {pages} {31303}
  (\bibinfo {year} {2013})}\BibitemShut {NoStop}%
\bibitem [{\citenamefont {Aladjidi}\ \emph {et~al.}(2022)\citenamefont
  {Aladjidi}, \citenamefont {Abuzarli}, \citenamefont {Brochier}, \citenamefont
  {Bienaimé}, \citenamefont {Picot}, \citenamefont {Bramati},\ and\
  \citenamefont {Glorieux}}]{aladjidi2022transit}%
  \BibitemOpen
  \bibfield  {author} {\bibinfo {author} {\bibfnamefont {T.}~\bibnamefont
  {Aladjidi}}, \bibinfo {author} {\bibfnamefont {M.}~\bibnamefont {Abuzarli}},
  \bibinfo {author} {\bibfnamefont {G.}~\bibnamefont {Brochier}}, \bibinfo
  {author} {\bibfnamefont {T.}~\bibnamefont {Bienaimé}}, \bibinfo {author}
  {\bibfnamefont {T.}~\bibnamefont {Picot}}, \bibinfo {author} {\bibfnamefont
  {A.}~\bibnamefont {Bramati}},\ and\ \bibinfo {author} {\bibfnamefont
  {Q.}~\bibnamefont {Glorieux}},\ }\href@noop {} {\bibinfo {title} {Transit
  effects for non-linear index measurement in hot atomic vapors}} (\bibinfo
  {year} {2022}),\ \Eprint {https://arxiv.org/abs/2202.05764} {arXiv:2202.05764
  [quant-ph]} \BibitemShut {NoStop}%
\bibitem [{\citenamefont {McCormick}\ \emph {et~al.}(2004)\citenamefont
  {McCormick}, \citenamefont {Solli}, \citenamefont {Chiao},\ and\
  \citenamefont {Hickmann}}]{mccormick2004saturable}%
  \BibitemOpen
  \bibfield  {author} {\bibinfo {author} {\bibfnamefont {C.}~\bibnamefont
  {McCormick}}, \bibinfo {author} {\bibfnamefont {D.}~\bibnamefont {Solli}},
  \bibinfo {author} {\bibfnamefont {R.}~\bibnamefont {Chiao}},\ and\ \bibinfo
  {author} {\bibfnamefont {J.}~\bibnamefont {Hickmann}},\ }\href@noop {}
  {\bibfield  {journal} {\bibinfo  {journal} {Physical Review A}\ }\textbf
  {\bibinfo {volume} {69}},\ \bibinfo {pages} {023804} (\bibinfo {year}
  {2004})}\BibitemShut {NoStop}%
\bibitem [{\citenamefont {Wang}\ \emph {et~al.}(2020)\citenamefont {Wang},
  \citenamefont {Yuan}, \citenamefont {Wang}, \citenamefont {Xiao},\ and\
  \citenamefont {Jia}}]{wang2020measurement}%
  \BibitemOpen
  \bibfield  {author} {\bibinfo {author} {\bibfnamefont {S.}~\bibnamefont
  {Wang}}, \bibinfo {author} {\bibfnamefont {J.}~\bibnamefont {Yuan}}, \bibinfo
  {author} {\bibfnamefont {L.}~\bibnamefont {Wang}}, \bibinfo {author}
  {\bibfnamefont {L.}~\bibnamefont {Xiao}},\ and\ \bibinfo {author}
  {\bibfnamefont {S.}~\bibnamefont {Jia}},\ }\href@noop {} {\bibfield
  {journal} {\bibinfo  {journal} {Optics Express}\ }\textbf {\bibinfo {volume}
  {28}},\ \bibinfo {pages} {38334} (\bibinfo {year} {2020})}\BibitemShut
  {NoStop}%
\bibitem [{\citenamefont {Hau}\ \emph {et~al.}(1999)\citenamefont {Hau},
  \citenamefont {Harris}, \citenamefont {Dutton},\ and\ \citenamefont
  {Behroozi}}]{hau1999light}%
  \BibitemOpen
  \bibfield  {author} {\bibinfo {author} {\bibfnamefont {L.~V.}\ \bibnamefont
  {Hau}}, \bibinfo {author} {\bibfnamefont {S.~E.}\ \bibnamefont {Harris}},
  \bibinfo {author} {\bibfnamefont {Z.}~\bibnamefont {Dutton}},\ and\ \bibinfo
  {author} {\bibfnamefont {C.~H.}\ \bibnamefont {Behroozi}},\ }\href@noop {}
  {\bibfield  {journal} {\bibinfo  {journal} {Nature}\ }\textbf {\bibinfo
  {volume} {397}},\ \bibinfo {pages} {594} (\bibinfo {year}
  {1999})}\BibitemShut {NoStop}%
\bibitem [{\citenamefont {Kr{\"u}ger}\ and\ \citenamefont
  {Scheel}(2019)}]{kruger2019interseries}%
  \BibitemOpen
  \bibfield  {author} {\bibinfo {author} {\bibfnamefont {S.~O.}\ \bibnamefont
  {Kr{\"u}ger}}\ and\ \bibinfo {author} {\bibfnamefont {S.}~\bibnamefont
  {Scheel}},\ }\href@noop {} {\bibfield  {journal} {\bibinfo  {journal}
  {Physical Review B}\ }\textbf {\bibinfo {volume} {100}},\ \bibinfo {pages}
  {085201} (\bibinfo {year} {2019})}\BibitemShut {NoStop}%
\bibitem [{\citenamefont {Rommel}\ \emph {et~al.}(2021)\citenamefont {Rommel},
  \citenamefont {Main}, \citenamefont {Kr{\"u}ger},\ and\ \citenamefont
  {Scheel}}]{rommel2021interseries}%
  \BibitemOpen
  \bibfield  {author} {\bibinfo {author} {\bibfnamefont {P.}~\bibnamefont
  {Rommel}}, \bibinfo {author} {\bibfnamefont {J.}~\bibnamefont {Main}},
  \bibinfo {author} {\bibfnamefont {S.~O.}\ \bibnamefont {Kr{\"u}ger}},\ and\
  \bibinfo {author} {\bibfnamefont {S.}~\bibnamefont {Scheel}},\ }\href@noop {}
  {\bibfield  {journal} {\bibinfo  {journal} {Physical Review B}\ }\textbf
  {\bibinfo {volume} {104}},\ \bibinfo {pages} {085204} (\bibinfo {year}
  {2021})}\BibitemShut {NoStop}%
\bibitem [{\citenamefont {Curwen}\ \emph {et~al.}(2019)\citenamefont {Curwen},
  \citenamefont {Reno},\ and\ \citenamefont {Williams}}]{curwen2019broadband}%
  \BibitemOpen
  \bibfield  {author} {\bibinfo {author} {\bibfnamefont {C.~A.}\ \bibnamefont
  {Curwen}}, \bibinfo {author} {\bibfnamefont {J.~L.}\ \bibnamefont {Reno}},\
  and\ \bibinfo {author} {\bibfnamefont {B.~S.}\ \bibnamefont {Williams}},\
  }\href@noop {} {\bibfield  {journal} {\bibinfo  {journal} {Nature Photonics}\
  }\textbf {\bibinfo {volume} {13}},\ \bibinfo {pages} {855} (\bibinfo {year}
  {2019})}\BibitemShut {NoStop}%
\bibitem [{\citenamefont {Wu}\ \emph {et~al.}(2021)\citenamefont {Wu},
  \citenamefont {Shen}, \citenamefont {Addamane}, \citenamefont {Reno},\ and\
  \citenamefont {Williams}}]{wu2021tunable}%
  \BibitemOpen
  \bibfield  {author} {\bibinfo {author} {\bibfnamefont {Y.}~\bibnamefont
  {Wu}}, \bibinfo {author} {\bibfnamefont {Y.}~\bibnamefont {Shen}}, \bibinfo
  {author} {\bibfnamefont {S.}~\bibnamefont {Addamane}}, \bibinfo {author}
  {\bibfnamefont {J.~L.}\ \bibnamefont {Reno}},\ and\ \bibinfo {author}
  {\bibfnamefont {B.~S.}\ \bibnamefont {Williams}},\ }\href@noop {} {\bibfield
  {journal} {\bibinfo  {journal} {Optics Express}\ }\textbf {\bibinfo {volume}
  {29}},\ \bibinfo {pages} {34695} (\bibinfo {year} {2021})}\BibitemShut
  {NoStop}%
\bibitem [{\citenamefont {Orfanakis}\ \emph {et~al.}(2021)\citenamefont
  {Orfanakis}, \citenamefont {Rajendran}, \citenamefont {Ohadi}, \citenamefont
  {Zieli{\'n}ska-Raczy{\'n}ska}, \citenamefont {Czajkowski}, \citenamefont
  {Karpi{\'n}ski},\ and\ \citenamefont {Ziemkiewicz}}]{OrfanakisNano}%
  \BibitemOpen
  \bibfield  {author} {\bibinfo {author} {\bibfnamefont {K.}~\bibnamefont
  {Orfanakis}}, \bibinfo {author} {\bibfnamefont {S.}~\bibnamefont
  {Rajendran}}, \bibinfo {author} {\bibfnamefont {H.}~\bibnamefont {Ohadi}},
  \bibinfo {author} {\bibfnamefont {S.}~\bibnamefont
  {Zieli{\'n}ska-Raczy{\'n}ska}}, \bibinfo {author} {\bibfnamefont
  {G.}~\bibnamefont {Czajkowski}}, \bibinfo {author} {\bibfnamefont
  {K.}~\bibnamefont {Karpi{\'n}ski}},\ and\ \bibinfo {author} {\bibfnamefont
  {D.}~\bibnamefont {Ziemkiewicz}},\ }\href@noop {} {\bibfield  {journal}
  {\bibinfo  {journal} {Physical Review B}\ }\textbf {\bibinfo {volume}
  {103}},\ \bibinfo {pages} {245426} (\bibinfo {year} {2021})}\BibitemShut
  {NoStop}%
\bibitem [{\citenamefont {Heck{\"o}tter}\ \emph {et~al.}(2018)\citenamefont
  {Heck{\"o}tter}, \citenamefont {Freitag}, \citenamefont {Fr{\"o}hlich},
  \citenamefont {A{\ss}mann}, \citenamefont {Bayer}, \citenamefont
  {Gr{\"u}nwald}, \citenamefont {Sch{\"o}ne}, \citenamefont {Semkat},
  \citenamefont {Stolz},\ and\ \citenamefont {Scheel}}]{heckotter2018rydberg}%
  \BibitemOpen
  \bibfield  {author} {\bibinfo {author} {\bibfnamefont {J.}~\bibnamefont
  {Heck{\"o}tter}}, \bibinfo {author} {\bibfnamefont {M.}~\bibnamefont
  {Freitag}}, \bibinfo {author} {\bibfnamefont {D.}~\bibnamefont
  {Fr{\"o}hlich}}, \bibinfo {author} {\bibfnamefont {M.}~\bibnamefont
  {A{\ss}mann}}, \bibinfo {author} {\bibfnamefont {M.}~\bibnamefont {Bayer}},
  \bibinfo {author} {\bibfnamefont {P.}~\bibnamefont {Gr{\"u}nwald}}, \bibinfo
  {author} {\bibfnamefont {F.}~\bibnamefont {Sch{\"o}ne}}, \bibinfo {author}
  {\bibfnamefont {D.}~\bibnamefont {Semkat}}, \bibinfo {author} {\bibfnamefont
  {H.}~\bibnamefont {Stolz}},\ and\ \bibinfo {author} {\bibfnamefont
  {S.}~\bibnamefont {Scheel}},\ }\href@noop {} {\bibfield  {journal} {\bibinfo
  {journal} {Physical Review Letters}\ }\textbf {\bibinfo {volume} {121}},\
  \bibinfo {pages} {097401} (\bibinfo {year} {2018})}\BibitemShut {NoStop}%
\end{thebibliography}%

\section{Supplementary Information}

\section{Experimental details}

\subsection{Sample and laser source}

The sample is a natural $\mathrm{Cu_2O}$ crystal procured commercially. It is $50~\si{\micro\meter}$ thick, highly polished on both sides, oriented so that light propagates along the [001] axis and held strain-free in a $4~\si{\kelvin}$ microscopy cryostat in transmission mode. We took great care to ensure a good thermalization between the sample and the holder; the best result was obtained using a small amount of vacuum grease in one sample corner.

The yellow laser is an all-automated doubled OPO system (C-Wave from Hubner Photonics) providing a CW beam (spectral width $\lesssim 1~\si{\mega\hertz}$) tunable to almost any wavelength within $450-650~\si{\nano\meter}$. The beam is focused on the sample with a waist diameter of about $400~\si{\micro\meter}$. We frequency-lock the laser onto a high-precision wavemeter to stabilize the wavelength at any setpoint with $\pm 30~\si{\mega\hertz}$ accuracy. Additionally, the laser intensity is locked to an arbitrary value with a feedback loop onto an acousto-optic modulator (AOM). This reduces the intensity fluctuations to $\lesssim 1\%$ of the mean value and allows to modulate the laser beam, thus avoiding any heating. We use $10~\si{\milli\second}$ pulses with a $10\%$ duty cycle.

A computer-controlled Red Pitaya FPGA board is used as a fast analog I/O card, as a PID lock box, and as a digital I/O card: it acquires the photodiodes signals, stabilizes the laser intensity and controls the flip mirror as well as several home-made mechanical shutters. The main computer orchestrates a fully-automated measurement protocol, including quality assessment routines, with which we acquired all the data presented in the main text.

\subsection{Phase front imaging}

Our off-axis interferometry imaging technique uses a modified Mach–Zehnder interferometer: one arm is focused on the sample plane and imaged on the camera, while the other is expanded with a divergent lens and used as a quasi-flat reference. The sample image and the reference beam arrive on the CMOS chip at an angle so that high-contrast fringes containing the phase profile information are present in the picture. To optimize the spatial resolution we chose an inter-fringe distance of about 10 pixels ($\sim 25~\si{\micro\meter}$) on the camera. The actual phase picture is then retrieved numerically with a Fourier transform algorithm (described below). The home-made mechanical shutters allow to switch between imaging the interference pattern, the intensity profile, or the reference beam. All three are recorded for each data point.

Note that, although the phase front deformation present in Figure 1 (b) of the main text corresponds to the onset of self-focusing, the $L=50~\si{\micro\meter}$ propagation distance between the input and the imaging plane is too small to observe any focusing: we experimentally confirmed that the small decrease in beam diameter seen between low and high power on the intensity pictures is due to nonlinear absorption. On the interference pictures, the nonlinear absorption changes the local contrast but not the fringes shape encapsulating the phase information.

\subsection{Numerical phase extraction}

The two incident beams on the camera are the reference beam (electric field $\vec{E_{R}}(\vec{r})=A(\vec{r})e^{i(\vec{k_{R}}.\vec{r} + \phi_R(\vec{r}))}\vec{u}$ in the camera plane) and the "signal" beam that passed through the sample (electric field $\vec{E_{S}}(\vec{r})=B(\vec{r})e^{i(\vec{k_{S}}.\vec{r} + \phi_S(\vec{r}))}\vec{u}$ in the camera plane). Thus the resulting intensity pattern on the camera can be expressed as $I(\vec{r})= I_0(\vec{r}) + A(\vec{r})B(\vec{r})\times e^{i[(\vec{k_{S}}-\vec{k_{R}}).\vec{r} + \phi_S(\vec{r}) - \phi_R(\vec{r})]} + c.c$. In the Fourier domain, the peak located at $\vec{k_{S}}-\vec{k_{R}} \neq 0$ contains the phase difference information $\phi_S(\vec{r}) - \phi_R(\vec{r})$. To recover this phase from the interference picture we therefore isolate this peak in the Fourier plane, shift it to a zero wave vector and compute the inverse Fourier transform. The spatial phase map is then the argument of the resulting complex array.

To automatize the numerical phase extraction we wrote a program able to find the relevant Fourier peaks. To do so, we implemented an iterative algorithm that binarized the FFT image with a descending threshold to locate the three dominant peaks (two correspond to $\vec{k}=\pm(\vec{k_{S}}-\vec{k_{R}})$ and one corresponds to $\vec{k}=0$). The area around $+(\vec{k_{S}}-\vec{k_{R}})$ is then cropped and copied to the center while the rest is zeroed out. The argument of the inverse 2D FFT is then stored and unwrapped to avoid $2\pi$ jumps. This procedure is equivalent to a demodulation operation.

As explained in the main text, two set of pictures are taken for each laser energy: one at very low power, used as a reference where the phase is mostly intensity-independent, and one at high power containing the intensity-dependent phase shift. Subtracting the two allows to remove all the repeatable, intensity-independent spatial fluctuations such as the slightly parabolic phase front of the reference beam, parasitic interference and optical aberrations. The remaining noise mainly originates from the air fluctuations, as they are not repeatable and, in our case, worsened by the presence of a cold cryostat exhaust. To reach a high phase resolution, we shrunk the interferometer around the cryostat ($\sim 10\sim 15$ cm arm length), used a triple isolation system (optical tubes along the beam path, thermal isolation foam around each arm and a plexiglass box encasing the whole interferometer) and integrated for well over the typical air fluctuation timescale. The reduced air volume and the good temperature stability enabled us to reach an almost perfect fringe contrast in spire of integrating over the remaining air fluctuation timescale, indicating that these fluctuations very low (shifting the interference by much less than the inter-fringe distance). 

The shutters and the camera are programmed to take both interference and intensity pictures. Therefore, each pixel of the intensity map can be linked to a pixel on the phase map: pixels of equal value in the intensity picture (within a tolerance) are selected to form a Boolean mask. This mask is used on the phase picture to compute the average phase for the selected intensity. Looping over the selected intensity, we find the intensity variation of the nonlinear phase shift $\Delta\phi(I)$ from a single pair of pictures. We systematically remove any artifact offset by setting the condition $\Delta\phi(0) = 0$. The error bars (in both direction) are computed from the standard deviation over the masked images.

\begin{figure*}[t!]
\begin{center}
  \includegraphics[width=0.9\linewidth]{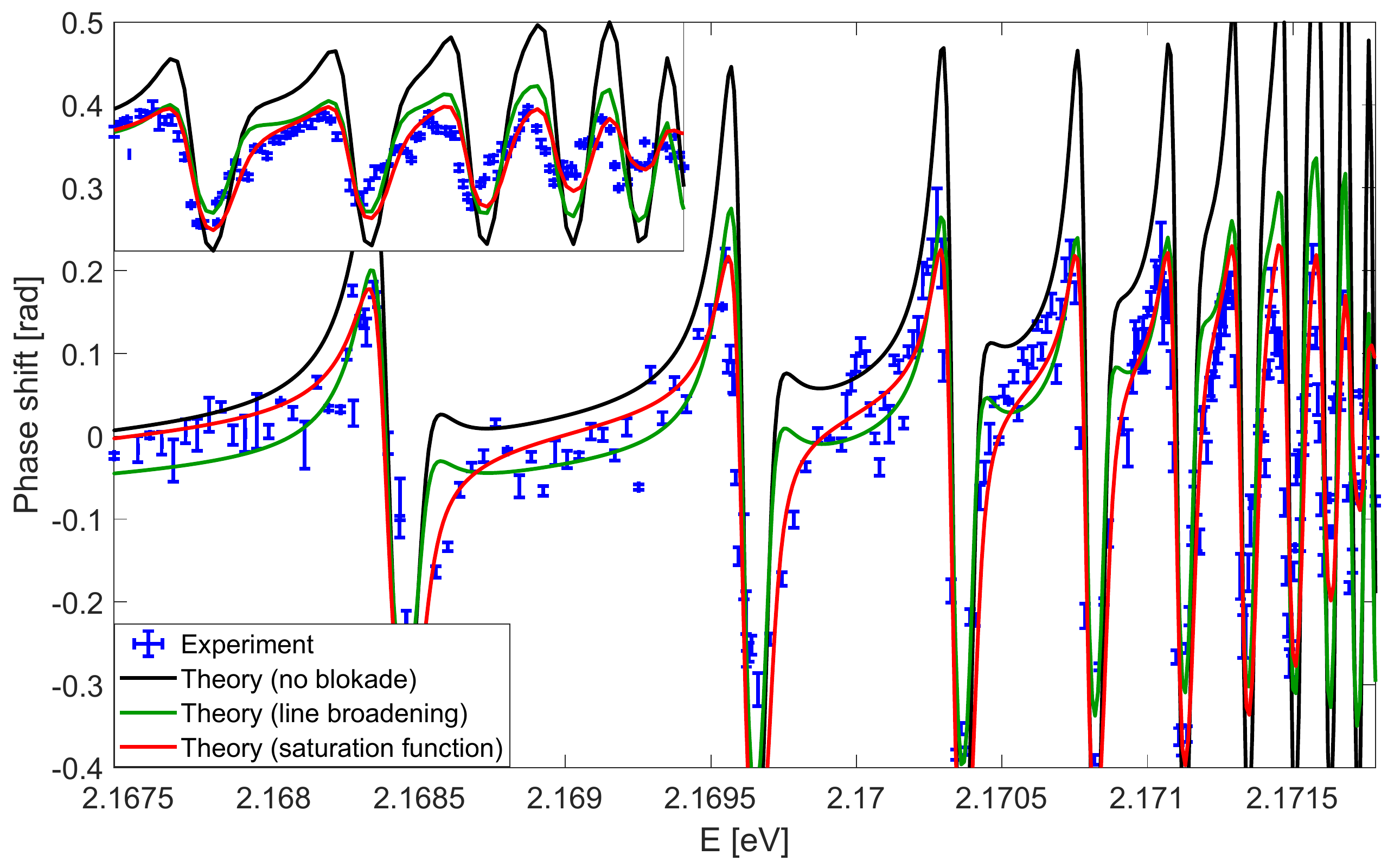}
  \caption{Comparison between different phenomenological ways to include the Rydberg interaction in the phase shift calculation: no interaction (dark green), collisional broadening (light green), saturable exciton density (red line). Including the saturation yields a better agreement with the experiment (blue datapoints).}
  \label{figSup-model}
\end{center}
\end{figure*}

\section{Theoretical model}

\subsection{Real Density Matrix Approach }
RDMA allows one to obtain analytical expressions for the optical functions of semiconductor crystals, including Rydberg excitons with arbitrarily high principal quantum numbers, also for the case of indirect interband transitions. Moreover, this method takes into account the effects of an anisotropic dispersion and coherence of the electron and the hole with the radiation field. Its general character consists in the fact that using a small number of well-known parameters (e.g., effective masses, gap energy, dielectric constant) one gets analytical expressions for optical functions of any crystal.

We consider the nonlinear response of a semiconductor slab to an electromagnetic wave characterized by the electric field vector,
\begin{equation}
{\bf E}=\textbf{E}_{i0}\exp( i\textbf{k}_0{\textbf{R}}-i\omega t),
\quad k_0=\omega/c,
\end{equation}
where \textbf{R} is the excitonic center-of-mass coordinate. In the RDMA approach, the bulk nonlinear response is described by a closed set of differential equations ("constitutive equations"): one for the coherent amplitude $Y({\bf r}_1, {\bf r}_2)$ representing the exciton density related to the interband transition, and for intraband transitions: one for the density matrix for electrons $C({\bf r}_1, {\bf r}_2)$ (assuming a non-degenerate conduction band), and one for the density matrix for the holes in the valence band, $ D({\bf r}_1, {\bf r}_2)$. Below we will use the notation
\begin{equation}
 Y({\bf r}_1, {\bf r}_2)=Y_{12},\quad\hbox{etc}.
\end{equation}
\noindent The constitutive equations have the following form: interband equation,
\begin{eqnarray}\label{interband1}
 & &{i}\hbar\partial_tY_{12}-H_{eh}Y_{12}=-{\bf M}{\bf E}({\bf R}_{12})\nonumber
\\
& &+{\bf E}_1{\bf M}_{0}C_{12}+{\bf E}_2{\bf M}_{0}D_{12}+{i}\hbar\left(\frac{\partial Y_{12}}{\partial t}\right)_{{\rm
irrev}},
\end{eqnarray}
conduction band equation,
\begin{eqnarray}
& &\label{conduction1}{i}\hbar\partial_tC_{12}+H_{ee}C_{12}={\bf M}_{0}({\bf E}_1 Y_{12}-{\bf E}_2Y^*_{21})\nonumber\\
& &+{i}\hbar\left(\frac{\partial C_{12}}{\partial t}\right)_{{\rm irrev}},\end{eqnarray} 
valence band equation,
\begin{eqnarray} &
&\label{valence1}{i}\hbar\partial_tD_{21}-H_{hh}D_{21}={\bf M}_{0}({\bf E}_2 Y_{12}-{\bf E}_1Y^*_{21})\nonumber\\
&&+ {i}\hbar\left(\frac{\partial D_{21}}{\partial t}\right)_{{\rm irrev}},
\end{eqnarray}
\noindent where the operator $H_{eh}$ is the effective mass Hamiltonian for the electron-hole pair, and
\begin{eqnarray}
&
&H_{ee}=-\frac{\hbar^2}{2m_e}(\hbox{\boldmath$\nabla$}_1^2-\hbox{\boldmath$\nabla$}_2^2),\nonumber
\\
&
&H_{hh}=-\frac{\hbar^2}{2m_{h}}(\hbox{\boldmath$\nabla$}_1^2-\hbox{\boldmath$\nabla$}_2^2).
\end{eqnarray}
\noindent ${\bf E}_{12}$ means that the wave electric field in the medium is taken in a middle point between ${\bf r}_1$ and ${\bf r}_2$. We take them at the center-of-mass
\begin{equation}\label{com}
{\bf R}={\bf R}_{12}=\frac{m_h{\bf r}_1+m_e{\bf r}_2}{m_h+m_e}.
\end{equation}
\noindent In the above formulas $m_e$ and $m_h$ are, respectively, the electron and the hole effective masses. The terms denoted as $(..)_{\rm irrev}$ describe the dissipation and radiation decay processes. The smeared-out transition dipole density ${\bf M}({\bf r})$ is related to the bilocality of the amplitude $Y$ and describes the quantum coherence between the macroscopic electromagnetic field and the inter-band transitions. The resulting coherent amplitude $Y_{12}$ determines the excitonic part of the polarization of the medium
\begin{eqnarray}\label{polarization1}
&&{\bf P}({\bf R},t)=2\int{\rm d}^3r\,\textbf{M}^*({\bf r})\hbox{Re}~Y({\bf R},{\bf r},t)\nonumber\\
&&=\int{\rm d}^3r\textbf{M}^*({\bf r})[Y({\bf R},{\bf r},t)+\hbox{c.c}],
\end{eqnarray}
\noindent where ${\bf r}={\bf r}_1-{\bf r}_2$ is the electron-hole relative coordinate.

The linear and nonlinear optical properties are obtained by solving iteratively the set of constitutive equations (\ref{interband1})-(\ref{valence1}), supplemented by the relation (\ref{polarization1}) and the corresponding Maxwell equation. The first step in the iteration consists of solving the linearized version of equation (\ref{interband1}) (i.e. by putting $C=D=0$) where we assume, as is usual, a relaxation time approximation for the irreversible part with a time constant denoted as $T_2$. Inserting the resulting linear amplitude $Y^{(1)}$, with the transition dipole density $\textbf{M}$ appropriate to $p$-excitons, into Eq. (\ref{polarization1}) gives the linear susceptibility

\begin{eqnarray}\label{polariton1}
&&\chi^{(1)}=\epsilon_b\sum\limits_{n=2}^N\frac{f_{n1}\Delta_{LT}}{E_{Tn}-E-i{\mit\Gamma}_n},
\end{eqnarray}
where  ${\mit\Gamma}_n=\hbar/T_{2n}$,  $E_{Tn}$ are the energies of the exciton resonances, $\Delta_{LT}$ is the
longitudinal-transversal splitting energy. The oscillator strengths are given by \begin{equation}\label{oscillatorforces}
f_{n1}=\frac{32(n^2-1)}{3n^5}\left[\frac{n(r_0+2a^*)}{2(r_0+na^*)}\right]^6,
\end{equation}
where $a^*=1.1$ nm is the Bohr radius and the coherence radius $r_0$ is treated as an adjustable parameter.

In the second iteration step, the linear amplitudes $Y^{(1)}$ are inserted into the source terms of the conduction and valence band equations (\ref{conduction1} - \ref{valence1}) and we are looking for stationary solutions. Assuming the relaxation time approximation one obtains the matrices $C(r), D(r)$ in the form
\begin{eqnarray}\label{CD}
&&C({\bf r})=-\frac{i}{\hbar}\left[\tau J_C({\bf r})-\tau J_C(r_0)+T_1f_{0e}({\bf r})J_C(r_0)\right],\nonumber\\
&&\\ & &D({\bf r})=-\frac{i}{\hbar}\left[\tau J_{V}({\bf r})-\tau J_{V}(r_0)+T_{1}f_{0h}({\bf r})J_{V}(r_0)\right].\nonumber
\end{eqnarray}
Here $J_{C,V}$ denote the source terms, $\tau$ is related to the relaxation of the quasi-particles distributions, $T_1$ stands for the interband recombination, and $f_{0e}$ is taken in the form
\begin{eqnarray}\label{edistribution}
&&{f}_{0e}(\textbf{r})={f_{0e}}(r,\theta,\phi)=\sqrt{\frac{\pi}{2}}\frac{r}{\lambda_{\hbox{\tiny th e}}}\nonumber\\
&&\times\sqrt{\frac{4\pi}{3}}\,Y_{10}(\theta,\phi)\,\exp\left(-\frac{r^2}{2\lambda_{\hbox{\tiny th e}}^2}\right),
\end{eqnarray}
where
\begin{eqnarray}\label{thermal}
&&\lambda_{\hbox{\tiny th e}}=\left(\frac{\hbar^2}{m_e k_B{\mathcal T}}\right)^{1/2}=\sqrt{\frac{2\mu}{m_e}}\sqrt{\frac{R^*}{k_B{\mathcal T}}}a^*,
\end{eqnarray}
is the so-called thermal length for the electrons at temperature $\mathcal T$. The same reasoning holds for holes, using the effective hole mass in Eq. (\ref{thermal}) to find the hole the thermal length.

The above expressions (\ref{CD}) are inserted in the nonlinear version of Eq. (\ref{interband1}) (i.e. by setting
$\textbf{ME}=0$), so as to obtain the equation for the third order coherent amplitude $Y^{(3)}_{12}$. The solutions of those equation, when substituted to Eq. (\ref{polarization1}), give the third-order contribution to the polarization, from which the susceptibility $\chi^{(3)}$ can be calculated. Assuming that $T_1\gg\tau$, we neglect the terms proportional to $\tau$ and obtain the following result
\begin{eqnarray}\label{chi3f}
&&\chi^{(3)}=-\chi^{(3)}_0\sum\limits_{nn'}\frac{{\mathcal F}_{nn'}{\mit\Gamma}_{n'}\,E_{Tn1}}{[(E_{Tn'1}-E)^2+{\mit\Gamma_{n'}}^2][E_{Tn1}^2-E^2-2iE{\mit\Gamma}_n]}\nonumber\\
&&{\mathcal F}_{jn}=\frac{(n'^2-1)(n^2-1)}{n'^5}\left(\frac{A({\mathcal T})}{n^\gamma}+\frac{B({\mathcal T})}{n^\beta}\right),\\
&&A({\mathcal T}=4\hbox {K})=4.53,\qquad B({\mathcal T}=4\hbox{K})=3.41,\nonumber\\
&&\gamma=1.8,\qquad \beta=1.62,\nonumber\\
&&\chi^{(3)}_0=\,0.6\times 10^{-11}\left[\frac{\hbox{m}^2}{\hbox{V}^2}\right]\nonumber,
\end{eqnarray}
where the summation is done over excitonic states with principal numbers $n$, $n'$, energies $E_{Tn}$, $E_{Tn'}$ and dissipation constants $\mit\Gamma_{n}$,$\mit\Gamma_{n'}$. Using the expressions (\ref{polariton1}) and (\ref{chi3f}), we can calculate the nonlinear absorption
\begin{eqnarray}\label{absorptionNL}
&&\alpha^{(3)}=\frac{\hbar\omega}{\hbar
c}\frac{1}{\sqrt{\epsilon_b}}\left(\hbox{Im}\,\chi^{(1)}+\vert
E_{\hbox{\tiny prop}}\vert^2\hbox{Im}\,\chi^{(3)}\right)
\end{eqnarray}
where $E_{prop}$ is the amplitude of the wave propagating in the crystal. It is obtained from the equation
\begin{eqnarray}\label{intensity}
&&\vert E_{prop}\vert^2=2\left|\frac{2}{1+\sqrt{\epsilon_b}}\right|^2\zeta P,
\end{eqnarray}
where $P$ is the laser power, and where $\zeta\approx 377\,\Omega$ is the impedance of free space. Having $\chi^{(3)}$, we are also able to determine the nonlinear index of refraction $\mathrm{n_2}$, defined as 
\begin{equation}\label{n2}
n_2=\frac{\hbox{Re}\,\{\chi^{(3)}\}}{c\,\epsilon_0n_0^2}
\end{equation}
where $n_0^2=1+\chi^{(1)}$. The total index of refraction is
\begin{eqnarray}
n^2=\epsilon_b+\chi^{(1)}+\vert
E_{\hbox{\tiny prop}}\vert^2\chi^{(3)}.
\end{eqnarray}
The phase shift is calculated from
\begin{equation}
\Delta \phi = \frac{\omega L}{c}\left[n(I)-n(0)\right],
\end{equation}
where $n(I)$ is the total refraction index obtained for average intensity $I$ inside the crystal and $L$ is the crystal length.

\subsection{Rydberg blockade inclusion}

The inclusion of the saturation induced by the Rydberg blockade is of particular importance; an extensive theoretical review of the effect of the Rydberg interaction potential on the nonlinear properties of Cu$_2$O is given in \cite{walther2020plasma}. While the calculations presented here are initially derived under the assumption that the saturation intensity is not reached, we propose two approaches to including the blockade effect. First, one can recall that in the blockade radius of one excitons, a second one cannot be created due to the energy shift of the excitonic lines. In a dynamic environment where the exciton-exciton distances are constantly changing, the time-variable shift results in an overall broadening of the excitonic line. Thus, one can use a fit function
\begin{equation}\label{eq_broadening}
\Gamma_n' = \Gamma_n + 2.1 \cdot 10^{-2}n^4P,
\end{equation}  
where $\Gamma_n$ is the unmodified linewidth of the state $n$ and $P$ is the input power. The constant is fitted to the particular geometry used in experiment. In the work~\cite{OrfanakisNano}, it has been shown that in Cu$_2$O nanoparticles, the additional confinement potential produces a shift of the excitonic energy levels. Due to randomness of the particle size, the resulting shift is also random; the overlap of shifted lines produces one, wider linewidth. This energy shift has been estimated to be proportional to $d^{-2}$, where $d$ is the nanoparticle diameter. In the same manner, we propose here that the blockade volumes of neighbouring excitons serve as a confinement potential for any given exciton, shifting it energy level by a random amount. For any given (power-dependent) density of excitons, the mean distance $d$ between blockade scales linearly with excitonic radius, and thus is proportional to $n^2$. Therefore, the overall broadening is proportional to $n^4$, as indicated in Eq. (\ref{eq_broadening}). Since the values of $\Gamma_n$ scale as $n^{-3}$, the relative broadening (and thus overall reduction of peak size) is proportional to $n^{7}$.

The second approach, guided by our experimental observations, is to use the saturable function as described in the main article. This approach is motivated by the fact that the Rydberg blockade effect is usually included in the oscillator strengths \cite{kazimierczuk2014giant} and thus it reduces the line height directly, instead of an indirect reduction through broadening. A comparison of the two methods is shown on the Fig.~\ref{figSup-model}. One can see that both approaches provide a good match to the measured data, but the broadening method introduces some distortions to the line shape that are not visible in the experiment. Finally, one can see that the results obtained without the inclusion of blockade effect overestimate the phase shift roughly by a factor of 2. Interestingly, while the saturation method produces an overall better match to the experimental data, there is some discrepancy in the nonlinear index spectrum on the Fig.~\ref{figSup-model}, especially for the highest states $n=13-14$. The two above methods exhibit slightly different scaling with incident light intensity/power. Since $\mathrm{n_2}$ is proportional to the initial slope $\partial \phi/\partial I$, a combined approach of simultaneous oscillator strength reduction and broadening might possibly provide the best fit to the data. This is also connected with the fact that the bleaching can be attributed to two distinct effects: Rydberg blockade and band gap reduction by electron-hole plasma~\cite{walther2020plasma,heckotter2018rydberg}.

\end{document}